\documentclass[final,1p,times]{elsarticle}

\usepackage{amssymb}
\usepackage{amsmath}

\usepackage{tikz}
\usepackage{scalerel}
\usepackage{pgfplots}
\usepackage{graphicx}
\usepackage{dcolumn}
\usepackage{bm}
\usepackage{esint}
\usepackage{color}
\usepackage{dsfont}
\usepackage{xurl}
\usepackage{hyperref}

\begin{document}

\begin{frontmatter}



\title{On finite-temperature Fredholm determinants}


\author[inst1]{Oleksandr Gamayun}

\affiliation[inst1]{organization={London Institute for Mathematical Sciences},
            addressline={Royal Institution, 21 Albemarle St}, 
            city={London},
            postcode={W1S 4BS},
            country={UK}}

\author[inst2]{Yuri Zhuravlev}

\affiliation[inst2]{organization={ Bogolyubov Institute for Theoretical Physics},
            city={Kyiv},
            addressline={14-b Metrolohichna St},
            postcode={03143}, 
            country={Ukraine}}

\begin{abstract}
We consider finite-temperature deformation of the sine kernel Fredholm determinants acting on the closed contours. 
These types of expressions usually appear as static two-point correlation functions in the models of free fermions
and can be equivalently presented in terms of Toeplitz determinants. The corresponding symbol, or the phase shift, is related to the temperature weight. 
We present an elementary way to obtain large-distance asymptotic behavior even when the phase shift has a non-zero winding number. 
It is done by deforming the original kernel to the so-called effective form factors kernel that has a completely solvable matrix Riemann-Hilbert problem. 
This allows us to find explicitly the resolvent and address the subleading corrections. We recover Szeg{\H o}, Hartwig and Fisher, and Borodin-Okounkov
asymptotic formulas.
\end{abstract}


\end{frontmatter}


\section{Introduction and motivation}
\label{secIntro}

Quantum integrable systems attract much attention among researchers due to the possibility of precisely addressing non-perturbative physical phenomena ranging from AdS/CFT  to condensed matter \cite{Beisert2011,ESSLER2005, Guan2022}. The exact evaluation of the correlation function even in the integrable system remains a challenge because of the necessity to perform summation over the form factors \cite{Korepin1993, DeNardis2022}. At zero temperatures, one can use numerical summation \cite{Caux_2009} or settle with the effective field theory methods \cite{Giamarchi_2003}. For non-zero temperatures, or more generically for finite entropy states, more sophisticated methods were developed: the quantum transfer matrix approach \cite{10.21468/SciPostPhysLectNotes.16} and related thermal form factors series \cite{Dugave_2013,Dugave_2014,Ghmann_2017}; Leclair-Mussardo approach \cite{LeClair1999,SALEUR2000602,Pozsgay2018}  and its generalizations in terms of the thermodynamic form factor program \cite{CortsCubero2019,CortesCubero2020}, the quench action \cite{Caux_2016},  and the ballistic fluctuation theory \cite{Doyon2019} among many others. 

At large coupling constants, closed expressions exist for the correlation functions of integrable systems in terms of Fredholm determinants of the generalized sine kernels \cite{Korepin1993}. 
Their asymptotic behavior can be computed via the non-linear steepest descend analysis of the corresponding Riemann-Hilbert problem \cite{Deift1993,Deift1997}. 
This method is quite technically involved\cite{KKKThesis}, therefore in \cite{10.21468/SciPostPhys.10.3.070,10.21468/SciPostPhysCore.5.1.006} we introduced a heuristic method of \textit{effective form factors } to extract large time and long distance asymptotic of two-point functions in XY-spin chain \cite{10.21468/SciPostPhys.10.3.070}, the system of one-dimensional anyons \cite{Zhuravlev2022}, the mobile impurity \cite{Gamayun2022,Gamayun2023}, and Hubbard model \cite{Gamayun2023sp}. 
The method's main idea is to extract the asymptotic behavior not from the thermodynamic expression of the Fredholm determinants but rather from the scaling analysis with the system size of the appropriately modified form factor series.  We have checked the validity of this approach numerically, but the connection with the non-linear steepest descend method remained elusive. 
In this paper, we partially solve this problem. More specifically, we address the static (equal time) case for which the typical two-point function can be expressed as a Fredholm determinant acting on $L_2(\mathds{S}^1)$ (or $\mathds{S}^1$ for shortness) 
\begin{equation}\label{toeplitzdef}
    T_x[\theta] = \det_{ \mathds{S}^1}\left(1+\hat{S}\right)
\end{equation}
here by $\mathds{S}^1$ we mean a set $\{q\in \mathds{C}, |q|=1 \}$, an action of the operator $\hat{S}$ is specified by its kernel $S(q,p)$ 
\begin{equation}
    (\hat{S} f)(q) = \int\limits_{\mathds{S}^1} S(q,p)f(p)dp= \int\limits_{|p|=1}S(q,p)f(p)dp.
\end{equation}
The kernel differs from the traditional sine kernel by a prefactor given by the weight function $\theta(q)$, specifically  
\begin{equation}
S(q,p) = \frac{\theta(q)}{2\pi i}
    \frac{e_+(p)e_-(q)-e_+(q)e_-(p)}{p-q},\qquad e_\pm (q) = q^{\pm x/2}. 
\end{equation}
The Fredholm determinant is understood as the standard Fredholm series 
\begin{equation}
        T_x[\theta] =  1 + \sum\limits_{n=1}^\infty \frac{1}{n!} \int\limits_{(\mathds{S}^1)^n}\det(S(q_j,q_k))_{j,k=1}^n \prod\limits_{j=1}^n dq_j.
\end{equation}
The other 
definitions and the effective ways to compute Fredholm determinants can be found in Ref. \cite{Bornemann_2009}.

Analogously, the continuous counterpart is a \textit{finite-temperature} sine kernel, acting on the real line ($L_2(\mathds{R})$). 
\begin{equation}\label{toeplitzdef1}
    T_x[\theta] = \det_{ \mathds{R}}\left(1+\hat{S}\right),\qquad S(q,p) = \frac{\theta(q)}{2\pi }
    \frac{\sin(x (p-q))}{p-q}. 
\end{equation}
Originally, this type of kernel arose in the studies of one-dimensional Bose gas for the correlation functions of the impenetrable bosons \cite{Korepin1993}. 
In that case, the prefactor $\theta(q)$ is proportional to the Fermi weight $\theta(q) \sim (e^{(q^2-\mu)/T}+1)^{-1}$, where $T$ is the temperature and $\mu$ is a chemical potential. 
We will not fix the specific form of $\theta(q)$ but rather discuss asymptotic for the quite generic class of smooth functions decaying fast enough at the infinities. 
Recently there was a lot of interest in exploring the properties of \textit{finite-temperature} deformations of the sine kernel determinants \cite{sk1,Claeys2024}. 
Partially, it is connected with the fact that the corresponding Riemann-Hilbert problem turns out to be much simpler than in the \textit{zero-temperature} case when the contour degenerates to the set of arcs or intervals \cite{Slavnov_2010}.  

Essentially, since $\theta(q)$ decays fast enough at infinity, the case of $ \mathds{R}$ is completely equivalent to $ \mathds{S}^1$ so we will focus mainly on considering \eqref{toeplitzdef}. The answer for the continuous kernels can be obtained by introducing the radius of $\mathds{S}^1$ and then sending it to infinity with the appropriate rescating of $x$. We are not going to discuss the possible peculiarities of taking such a limit, we just mention that the results for the asymptotic behavior in the mobile impurity problems \cite{Gamayun2022} suggest that everything seems to work without any subtleties.

The great advantage of the  $ \mathds{S}^1$  case is that the Fredholm determinant \eqref{toeplitzdef} is nothing but a Toeplitz determinant (see \cite{Deift1999} or Sec. 5 in \cite{10.21468/SciPostPhys.10.3.070})
\begin{equation}
     T_x[\theta] = \det_{0\le n,m\le x-1} c_{n-m},\qquad \quad c_k = \frac{1}{2\pi i}\int\limits_{\mathds{S}^1}\frac{dq}{q^{k+1}}e^{2\pi i \nu(q)}.
\end{equation}
with 
\begin{equation}\label{nuTheta}
    e^{2\pi i \nu(q)} = 1+ \theta(q). 
\end{equation}
Throughout the rest of the paper, we assume the connection \eqref{nuTheta} and will treat $\nu(q)$ and $\theta(q)$ on equal footing. 
We assume that the phase shift $\nu(q)$ is a smooth function with a possible non-zero winding number around $\mathds{S}^1$. 
 The large $x$ asymptotic of the Toeplitz determinant for these types of symbols is known to be given by Szeg{\H o} formulas \cite{Szeg1915,grenander2001toeplitz} and its generalization by Hartwig and Fisher  \cite{Hartwig_1969,Bttcher2006}. The subleading corrections can be found in a form that is widely referred to as the Borodin-Okounkov formula \cite{Borodin2000}. 
 By trying to reproduce the effective form factor asymptotic for the determinant \eqref{toeplitzdef} we re-derive these results in elementary fashion. 
 It turns out that the effective form factors lead to such a deformation of the kernel that the corresponding  Riemann-Hilbert problem can be solved exactly. 
 As a sequence of that the resolvent is known explicitly and the subleading corrections can be reproduced systematically and presented in a closed form. 

The paper is organized as follows: in the next section \eqref{secS1}, we introduce the main notations and explain how the Riemann-Hilbert problem can be solved explicitly for the deformed kernels. 
In Sec. \eqref{Sec:var} we proceed to the evaluation of the Fredholm determinant by the means of computing its variation with respect to the phase shift. In Sec. \eqref{SubLead} we consider the subleading corrections, and in Sec. \eqref{Sec:effective} the connection to the effective form factors is thoroughly discussed. The last section contains concluding remarks and possible future directions. 
Various technical aspects and alternative formulations can be found in the Appendices. 

\section{Riemann-Hilbert Problem}
\label{secS1}

\begin{figure}[h]
	\centering
	\includegraphics[width=0.7\linewidth]{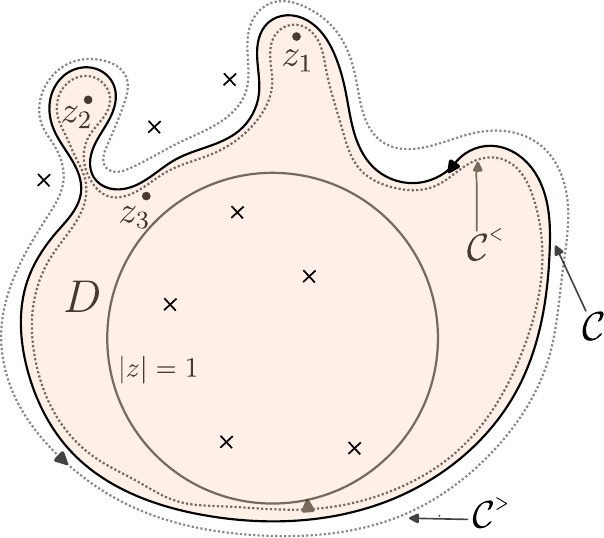}
	\caption{Schematic illustration of the contour  $\mathcal{C}$ in definition of $\tau_\mathcal{C}$ in Eq. \eqref{tauC} for $\mathfrak{n}=-3$. In this case we  have to encircle points $z_1$, $z_2$ and $z_3$, which lie outside the unit circle ($|z_k|>1$ ) and satisfy $1+\theta(z_k)=0$. The smooth contraction of the contour $\mathcal{C}$ back to $\mathds{S}^1$ ($|z|=1$) should not cross the poles of $\theta(q) $ schematically presented with cross marks. The dotted lines show the contours $\mathcal{C}^{\scaleto{>}{3pt}}$ and $\mathcal{C}^{\scaleto{<}{3pt}}$ that lie slightly above and below $\mathcal{C}$ correspondingly. The colored area inside $\mathcal{C}$ is denoted as $D$.}
	\label{fig1}
\end{figure}

In this section, we introduce main notations and give the formula for the resolvent of the kernel \eqref{toeplitzdef}. 
Let $\theta(q)$ in \eqref{toeplitzdef} be a meromorphic function in $\mathds{CP}^1$. We additionally assume that the 
function $\nu(q)$ in \eqref{nuTheta} doesn't have singularities on $ \mathds{S}^1$ , but might have  
a winding number $\mathfrak{n}$   
\begin{equation} 
    \mathfrak{n} = \oint_{ \mathds{S}^1} d\nu(q) \equiv \frac{1}{2\pi i} \oint_{ \mathds{S}^1}dq  \frac{\theta'(q)}{1+\theta(q)}.
\end{equation}
The winding number measures the difference between the number of zeros and the number of poles of $1+\theta(q)$ within $\mathds{S}^1$, or the difference between the number of poles and zero outside  $\mathds{S}^1$. 
Therefore, if $\mathfrak{n}<0$ there are at least $|\mathfrak{n}|$ zeroes outside of $\mathds{S}^1$. We order them by the absolute value $|z_1|>|z_2| \dots > |z_{-\mathfrak{n}}|>1$, $1+\theta(z_k) = 0$. 
For clarity, we assume that all zeroes are distinct even by their absolute value, but the proper limiting procedures can easily lift this restriction in the final answers. 
We want to deform the contour $\mathds{S}^1$ in \eqref{toeplitzdef} into $\mathcal{C}$ such that it will encircle all of the zeroes $\{z_k,\, k=1,\dots, -\mathfrak{n}\}$ but none of the poles of $\theta(q)$ (see Fig. \eqref{fig1}). 
By doing this the value of the Fredholm determinant remains unchanged, but the winding number of the new contour is zero 
\begin{equation}
    T_x[\theta] = \det_\mathcal{C}(1+\hat{S}),\qquad \oint_{\mathcal{C}} d\nu(q) =0.
\end{equation}
Similarly if $\mathfrak{n}>0$ we find zeroes  within $\mathds{S}^1$,  $1> |z_1|> |z_2| \dots > |z_{\mathfrak{n}}|$, $1+\theta(z_k) = 0$, and deform the  contour $\mathds{S}^1 \to \mathcal{C}$ to exclude them. 

Our next step is to deform the kernel. Namely, 
we introduce the deformed Fredholm determinant $\tau_{\mathcal{C}}[\nu]$ defined on the contour $\mathcal{C}$ 
\begin{equation}\label{tauC}
\tau_{\mathcal{C}}[\nu] = \det\limits_{\mathcal{C}} (1 + \hat{V}_{\mathcal{C}})
\end{equation}
with 
\begin{equation}\label{Vdef1}
 V_{\mathcal{C}}(q,p) =\frac{\sqrt{\theta(p)}\sqrt{\theta(q)}}{2\pi i}
    \frac{    \text{v}_+(p)    \text{v}_-(q)-    \text{v}_+(q)    \text{v}_-(p)}{p-q},
\end{equation}
\begin{equation}\label{Vdef2}
    \text{v}_- (q) = q^{-x/2},\qquad         \text{v}_+ (q) =     \text{v}_-(q) w_{\mathcal{C}}(q),\qquad w_{\mathcal{C}}(q)=q^{x}+ \int\limits_{\mathcal{C}^{\scaleto{<}{3pt}}}\frac{dk}{2\pi i} \frac{k^x }{k-q} \frac{\theta(k)}{1+\theta(k)}.
\end{equation}
Here the contour $\mathcal{C}^{\scaleto{<}{3pt}}$ lies slightly below $\mathcal{C}$ (to the left for the counterclockwise orientation, see Fig. \eqref{fig1}).  Notice that if we discard the last term in $V_{\mathcal{C}}$ we reproduce the kernel $\hat{S}$ up to conjugation with diagonal matrices that does not change the value of the determinant.  
The idea of such a deformation is that the resolvent for this kernel defined as 
\begin{equation}
    (1+ \hat{V}_{\mathcal{C}})(1-\hat{R})=1.
\end{equation}
can be found explicitly. It also has a similar (integrable) form 
\begin{equation}\label{r0}
     R(q,p) =  \frac{\sqrt{\theta(p)}\sqrt{\theta(q)}}{2\pi i}
    \frac{f_+(p)f_-(q)-f_+(q)f_-(p)}{p-q}.
\end{equation}
The functions $f_\pm$ and $\text{v}_{\pm}$ can be related by the integral equations 
\begin{equation}\label{int}
 (1+\hat{V}_{\mathcal{C}})\sqrt{\theta}f_\pm =\sqrt{\theta}     \text{v}_\pm,\qquad (1-\hat{R})\sqrt{\theta} \text{v}_\pm =\sqrt{\theta} f_\pm.
\end{equation}

These integral equations can be rewritten as a Riemann--Hilbert problem (RHP). To do this, we, following Slavnov \cite{Slavnov1996}, introduce the vector notations 
\begin{align}
\label{vectornotat}
    |V(q)\rangle &= \left(\begin{array}{c}
		    \text{v}_-(q) \\ -    \text{v}_+(q)
	\end{array}\right),\qquad \langle V(q)|  = (    \text{v}_+(q),    \text{v}_-(q)),\\
 |F(q)\rangle & = \left(\begin{array}{c}
		f_-(q) \\ -f_+(q)
	\end{array}\right),\qquad \langle F(q)|  = (f_+(q),f_-(q)),
\end{align}
and define the matrix function $\chi_{\mathcal{C}}(q)$
\begin{equation}\label{chiint}
    \chi_{\mathcal{C}}(q) = \mathds{1} - \int\limits_{\mathcal{C}} \frac{dk}{2\pi i} \frac{\theta(k) |F(k)\rangle\langle V(k)|}{k-q}.
\end{equation}
In these notations, the resolvent reads 
\begin{equation}\label{resolvent}
    R(p,q) = \frac{\sqrt{\theta(p)}\sqrt{\theta(q)}}{2\pi i}
    \frac{\langle F(p)| F(q)\rangle}{p-q}.
\end{equation}
One can immediately see that the integral equations \eqref{int} are equivalent to the $2\times 2$ matrix relation
\begin{equation}\label{fv}
    |F(q)\rangle=\chi_{\mathcal{C}}(q)|V(q)\rangle, \qquad q \in \mathcal{C}.
\end{equation}
Therefore finding $\chi_{\mathcal{C}}(q)$ is equivalent to finding the resolvent. This can be done by 
considering the jump of this function when $q$ crosses the contour $\mathcal{C}$, which is the essence of an RHP.  More specifically, 
let us denote 
 by $\chi_{\scaleto{>}{3pt}}$ the function analytic in $D$ s.t. $\partial D = \mathcal{C}$ (see Fig. \eqref{fig1}), and similarly 
 by  $\chi_{\scaleto{<}{3pt}}$ the function analytic in $\mathds{CP}^1 \setminus D$. Formally, they can be presented as
\begin{equation}\label{chibm}
      \chi_{\scaleto{>}{3pt}}(q) = \mathds{1} - \int\limits_{\mathcal{C}^{\scaleto{>}{3pt}}} \frac{dk}{2\pi i} \frac{\theta(k) |F(k)\rangle\langle V(k)|}{k-q},\qquad
       \chi_{\scaleto{<}{3pt}}(q) = \mathds{1} - \int\limits_{\mathcal{C}^{\scaleto{<}{3pt}}} \frac{dk}{2\pi i} \frac{\theta(k) |F(k)\rangle\langle V(k)|}{k-q}.
\end{equation}
It turns out that from the 
\begin{itemize}
    \item normalization condition 
    $\chi_{\scaleto{<}{3pt}}(q) \to\mathds{1} $ as $q\to \infty$,
    \item  and the jump condition for $q\in \mathcal{C}$ 
    \begin{equation}\label{jumpdef}
\chi_{\scaleto{<}{3pt}}(q)  =\chi_{\scaleto{>}{3pt}}(q)  J,\qquad 
J =  \mathds{1} + \theta(q) |V(q)\rangle \langle V(q)|,
\end{equation}
\end{itemize}
one can recover $\chi_{\mathcal{C}}$ explicitly due to the special form of the jump matrix $J$,
namely one can check by a direct computation that 
\begin{equation}\label{rhsol}
    \chi_{\mathcal{C}}(q)= 
 \left(\begin{array}{cc}
    	1&  b_{\mathcal{C}}(q)\\
    	0 & 1
    \end{array}\right)   \left(\begin{array}{cc}
    	e^{-\Omega_{\mathcal{C}}(q)} & 0\\
    	0 & e^{\Omega_{\mathcal{C}}(q)} 
    \end{array}\right)
    \left(\begin{array}{cc}
		1 & 0 \\
		\varphi_{\mathcal{C}}(q)& 1
	\end{array}\right),
\end{equation}
with
\begin{align}\nonumber
    \varphi_{\mathcal{C}} (q)& = \int\limits_{\mathcal{C}} \frac{dk}{2\pi i} \frac{k^x \theta(k) }{(k-q)(1+\theta(k))},\qquad 
       \Omega_{\mathcal{C}} (q) = \int\limits_{\mathcal{C}} \frac{\nu(k)dk}{k-q},\\ \label{bpint}
        b_{\mathcal{C}}(q) &= - \int\limits_{\mathcal{C}} \frac{dk}{2\pi i} \frac{k^{-x}\theta(k) e^{- \Omega_{\scaleto{>}{3pt}}(k)-\Omega_{\scaleto{<}{3pt}}(k)}}{k-q}.
\end{align}
The limiting values of $  \Omega_{\mathcal{C}} (q) $ for $q\in \mathcal{C}$ can be associated with the functions $\Omega_{\scaleto{>}{3pt}}$ and $\Omega_{\scaleto{<}{3pt}}$ that are defined similar to \eqref{chibm} 
\begin{equation}\label{OmegaPM}
    \Omega_{\scaleto{>}{3pt}}(q) = \int\limits_{\mathcal{C}^{\scaleto{>}{3pt}}} \frac{\nu(k)dk}{k-q},\qquad 
    \Omega_{\scaleto{<}{3pt}}(q) = \int\limits_{\mathcal{C}^{\scaleto{<}{3pt}}}  \frac{\nu(k)dk}{k-q}.
\end{equation}
These functions solve a scalar RHP on the contour $\mathcal{C}$ with the jump $2\pi i \nu$ i.e. 
\begin{equation}\label{scalarRHP}
        \Omega_{\scaleto{>}{3pt}}(q) -    \Omega_{\scaleto{<}{3pt}}(q)  = 2\pi i \nu(q). 
\end{equation}
Similar properties can be formulated for the functions $\varphi_{\mathcal{C}}$, $b_{\mathcal{C}}$ with the corresponding jumps, which turn the verification that \eqref{rhsol} solves the matrix RHP \eqref{jumpdef} into a straightforward exercise. 
To recover $f_\pm$ from the relation \eqref{fv} we can use either $\chi_{\scaleto{>}{3pt}}$ or $\chi_{\scaleto{<}{3pt}}$. In particular, we can explicitly present 
\begin{equation}\label{fmfp}
    f_-(q) = e^{-\Omega_{\scaleto{>}{3pt}}(q) }q^{-x/2} - b_{\scaleto{>}{3pt}}(q)  e^{\Omega_{\scaleto{<}{3pt}}(q)}q^{x/2},\qquad 
    f_+(q) =  e^{\Omega_{\scaleto{<}{3pt}}(q)}q^{x/2}.
\end{equation}
Notice that we have to specify the subscripts $>$ and $<$ when $q \in \mathcal{C}$;
for $q$ away from contour we will always use the expression analytically continued to the corresponding domain.

\section{Variational formulas}\label{Sec:var}

In this section using the exact form of the resolvent, we evaluate the tau function \eqref{tauC}. 
To do this we consider the variation $\nu \to \nu+\delta \nu$ such that the winding number around the contour $\mathcal{C}$ remains zero. 
The variation of the determinant can be computed as 
\begin{equation}
    \delta \ln \tau_{\mathcal{C}}[\nu] = {\rm Tr} (1 -\hat{R})\delta V_{\mathcal{C}}.
\end{equation}
The variation of the kernel reads
\begin{equation}\label{var1}
    \delta V_{\mathcal{C}}(q,p) = V_{\mathcal{C}}(q,p)  \frac{\delta\theta(p)}{2\theta(p)} +   \frac{\delta\theta(q)}{2\theta(q)} V_{\mathcal{C}}(q,p) 
    +  \frac{\sqrt{\theta(p)\theta(q)}}{2\pi i} q^{-x/2}p^{-x/2}\int\limits_{\mathcal{C}^{\scaleto{<}{3pt}}} \frac{dk}{2\pi i}
    \frac{\delta \theta(k)}{(1+\theta(k))^2}  \frac{k^x}{(k-q)(k-p)}.
\end{equation}
The trace of the first two terms can be reduced to the computation of the resolvent on the diagonal. Indeed, using $     (1 - \hat{R})\hat{V}_{\mathcal{C}} = \hat{R}$ we obtain
 \begin{equation}\label{var2}
     \delta \ln \tau_{\mathcal{C}}[\nu] 
= 
     \int\limits_\mathcal{C} dk\delta \theta(k) \frac{R(k,k)}{\theta(k)} 
   +   \int\limits_{\mathcal{C}^{{\scaleto{<}{3pt}}}} dk\delta \theta(k) \frac{k^x }{(1+\theta(k))^2} \frac{M(k,k)}{2\pi i} ,
 \end{equation}
where we have denoted by $M(k_1,k_2)$ the following combination 
\begin{equation}\label{Mdef}
   M(k_1,k_2) \equiv \int\limits_{\mathcal{C}} \frac{dq dp}{2\pi i}\frac{\sqrt{\theta(q)}q^{-x/2}}{k_1-q} (1-\hat{R})(q,p)\frac{\sqrt{\theta(p)}p^{-x/2}}{k_2-p}.
\end{equation}
One can easily express these integrals in terms of the solutions of the RHP (see \eqref{lemmaAPP} for details). 
\begin{equation}
   M(k_1,k_2) =  \frac{\chi_{22}(k_1)\chi_{12}(k_2)-\chi_{12}(k_1)\chi_{22}(k_2)}{k_1-k_2} 
\end{equation}
or using explicit presentation \eqref{rhsol}  
\begin{equation}\label{fromL2}
	M(k_1,k_2) = - e^{\Omega_{\mathcal{C}}(k_1)+\Omega_{\mathcal{C}}(k_2)} \frac{b_{\mathcal{C}}(k_1)-b_{\mathcal{C}}(k_2)}{k_1-k_2}.
\end{equation}
Here we should use either $\chi_{\scaleto{>}{3pt}}$ or $\chi_{\scaleto{<}{3pt}}$, and correspondingly $b_{\mathcal{C}}$ depending on the position of $k_1$ and $k_2$. In particular, for  $k_1=k_2=k \in \mathcal{C}^{\scaleto{<}{3pt}}$ using L'Hôpital's rule we obtain
$
    M(k,k) = - e^{2\Omega_{\scaleto{>}{3pt}}(k)} \partial_k b_{\scaleto{>}{3pt}}(k).
$ 

The resolvent at the diagonal can be easily evaluated with the explicit form of the $f_\pm$ (see Eqs. \eqref{fmfp}). 
It is convenient to present it in a form that can be continued into the domain $D$ so that integration over $\mathcal{C}$ can be replaced by the integration over $\mathcal{C}^{\scaleto{<}{3pt}}$, namely
\begin{equation}
      R(q,q) = \frac{\theta(q)}{2\pi i(1+\theta(q))} \left(x/q+ \partial_q \Omega_{\scaleto{>}{3pt}} + \partial_q \Omega_{\scaleto{<}{3pt}}\right)+ \frac{\theta(q)e^{2\Omega_{\scaleto{>}{3pt}}(q)}q^x}{2\pi i(1+\theta(q))^2}\partial_q b_{\scaleto{>}{3pt}}(q).
\end{equation}
Altogether the variation reads
 \begin{equation}\label{dlogtau}
     \delta \ln \tau_{\mathcal{C}}[\nu] 
= \int\limits_{\mathcal{C}} dq \delta \nu(q) \left(
    \frac{x}{q} + \int\limits_{\mathcal{C}^{\scaleto{<}{3pt}}}  \frac{\nu(k)dk}{(k-q)^2} +\int\limits_{\mathcal{C}^{\scaleto{>}{3pt}}}  \frac{\nu(k)dk}{(k-q)^2} 
    \right). 
     \end{equation}
Here we have replaced the variation in $\theta$ to the variation in $\nu$ using Eq. \eqref{nuTheta}. 
Taking into account that for $\nu(q) = 0$, $\tau_{\mathcal{C}}[\nu] =1$  we obtain 
\begin{equation}\label{tauCanswer}
 \boxed{\tau_{\mathcal{C}}[\nu] = \exp\left(x \int\limits_{\mathcal{C}} \frac{dq}{q}  \nu(q) 
 - \frac{1}{2}\int\limits_{\mathcal{C}} \int\limits_{\mathcal{C}} \left(\frac{\nu(k)-\nu(q)}{k-q}\right)^2 dk \, dq\right).}
\end{equation}
This formula gives the leading asymptotic for the Toeplitz determinant of the symbol with the properties indicated at the beginning of the Sec. \eqref{secS1}.

Let us discuss equivalent ways to present this formula and how certain known cases can be recovered from it. First, we notice that with the functions $\Omega_{\scaleto{>}{3pt}}$ and $\Omega_{\scaleto{<}{3pt}}$ from Eq. \eqref{OmegaPM} we can rewrite \begin{equation}\label{xi22}\int\limits_{\mathcal{C}} \frac{\nu(q) dq}{q}  = \int\limits_{\mathcal{C}} \frac{( \Omega_{\scaleto{>}{3pt}}(q) -    \Omega_{\scaleto{<}{3pt}}(q))dq}{2\pi i q} = \Omega_{\scaleto{>}{3pt}}(0). \end{equation}Here in the last integral we have shrinked the contour $\mathcal{C}$ into inside for $\Omega_{\scaleto{>}{3pt}}$ and outside for $\Omega_{\scaleto{<}{3pt}}$. Using integration by parts we can express $\tau_{\mathcal{C}}$ as 
\begin{equation}\label{tauCanswer2}
    \tau_{\mathcal{C}}[\nu] = \exp\left(x \Omega_{\scaleto{>}{3pt}}(0) + 
 \int\limits_{\mathcal{C}} \frac{dq}{2\pi i}\Omega'_{\scaleto{<}{3pt}}(q)\Omega_{\scaleto{>}{3pt}}(q)\right).
\end{equation}

For zero winding number we put $\mathcal{C}=\mathds{S}^1$ and present $\nu(q)$ via the Laurent series  $\nu(q) = \sum_{n \in \mathds{Z}} q^{n} \nu_n/(2\pi i)$. Then $\Omega_{\scaleto{>}{3pt}}(q) = \sum_{j\ge 0}\nu_j q^j$,
and 
 $\Omega_{\scaleto{<}{3pt}}(q) = -\sum_{j< 0}\nu_j q^j$ so 
Eq. \eqref{tauCanswer2} gives 
\begin{equation}
    \tau_{\mathcal{C}}[\nu]\Big|_{\mathcal{C}=\mathds{S}^1} =\exp\left(x \nu_0 + \sum\limits_{j=1}^{\infty} j \nu_j \nu_{-j}\right).
\end{equation}
which is nothing but the strong Szeg\H o formula for the asymptotic of the Toeplitz determinant. 
Notice that integrating by parts $\tau_{\mathcal{C}}$ in this case can be also presented as 
\begin{equation}
    \tau_{\mathcal{C}}[\nu]\Big|_{\mathcal{C}=\mathds{S}^1}= \exp\left(x \nu_0 + \int\limits_{\mathds{S}^1}\int\limits_{\mathds{S}^1} \nu'(k)\nu'(q) \ln|k-q|dkdq\right).
\end{equation}
The absolute value in the logarithm reflects that while integrating by parts the obtained expressions should be considered as principal value. 

Now let us turn to the case of a non-zero winding number of $\nu(q)$ w.r.t. $\mathds{S}^1$. We focus on the negative values $\mathfrak{n} = -n<0$. Then the contour $\mathcal{C}$ is chosen to encircle first $n$ roots of $1+\theta(q)$ lying outside the unit circle $z_1,z_2, \dots z_n$.  
For practical reasons, the integration over this contour can be somewhat cumbersome so we further transform the contour $\mathcal{C}$ in the r.h.s. of \eqref{tauCanswer} back to $\mathds{S}^1$. To do so we introduce the jump $\nu_{\mathfrak{n}}(q)$ that has zero winding number on $\mathds{S}^1$. This can be achieved by many ways but we choose the following modification
\begin{equation}\label{nudeltadef}
    \nu_{\mathfrak{n}}(q) = \nu(q)- \mathfrak{n} \frac{\ln(q) + i\pi}{2\pi i},\qquad \Rightarrow \qquad \int\limits_{\mathds{S}^1} \nu_{\mathfrak{n}}'(q) dq =0.
\end{equation}
Similar to Eq. \eqref{OmegaPM} we define the functions 
\begin{equation}\label{OmegaPM2}
    \omega_{\scaleto{>}{3pt}}(q) = \int\limits_{|k|=1+0} \frac{\nu_{\mathfrak{n}}(k)dk}{k-q},\qquad 
    \omega_{\scaleto{<}{3pt}}(q) = \int\limits_{|k|=1-0}  \frac{\nu_{\mathfrak{n}}(k)dk}{k-q},
\end{equation}
where the integration contours lie slightly above (below) the unit circle for $\omega_{\scaleto{>}{3pt}}$ $(\omega_{\scaleto{<}{3pt}})$. 
These functions  solve the scalar RHP with the modified jump $\omega_{\scaleto{>}{3pt}}(q)-\omega_{\scaleto{<}{3pt}}(q)=2\pi i\nu_{\mathfrak{n}}(q)$  (cf.  \eqref{scalarRHP}). 
By construction  $\omega_{\scaleto{<}{3pt}}$ $(\omega_{\scaleto{>}{3pt}})$ are analytic outside (inside) the unit circle. 
To connect these functions with $\Omega_{\scaleto{\lessgtr}{5pt}}$ we compare derivates of $\omega_{\scaleto{<}{3pt}}$ and $\Omega_{\scaleto{<}{3pt}}$ in the domain where both of them are analytic functions i.e. in $\mathds{CP}^1 \setminus D$. 
Integrating by parts we obtain 
\begin{equation}
	\omega_{\scaleto{<}{3pt}}'(q) = \int\limits_{|k|=1-0}  \frac{\nu'_{\mathfrak{n}}(k)dk}{k-q} = \int\limits_{|k|=1-0}  \frac{\nu'(k)dk}{k-q} + \frac{\mathfrak{n}}{q}.
\end{equation}
Computing  $	\Omega_{\scaleto{<}{3pt}}'(q) $ by the deformation the integration  contour $\mathcal{C}$ back to $\mathds{S}^1$ we have to account for the residues at points $z_k$, thus
\begin{equation}
	\Omega_{\scaleto{<}{3pt}}'(q) = \int\limits_{\mathcal{C}}  \frac{\nu'(k)dk}{k-q} = \int\limits_{|k|=1-0}  \frac{\nu'(k)dk}{k-q} -\sum\limits_{k=1}^n \frac{1}{z_k-q}.
\end{equation}
 This way, we conclude that 
 \begin{equation}\label{omegaomega1}
 e^{\omega_{\scaleto{<}{3pt}}(q)}=e^{\Omega_{\scaleto{<}{3pt}}(q)}\prod_{k=1}^n(1-z_k/q)
 \end{equation}
 The integration constant is chosen such that $\omega_{\scaleto{<}{3pt}}(q) \to 0 $ as $q \to \infty$. 
 The relation between $\omega_{\scaleto{>}{3pt}}(q)$ and $\Omega_{\scaleto{>}{3pt}}(q)$ can be found either directly as above, or analytically continuing  Eq. \eqref{omegaomega1} via the scalar RHPs 
   \begin{equation}\label{omegaomega2}
 	e^{\omega_{\scaleto{>}{3pt}}(q)}=e^{\Omega_{\scaleto{>}{3pt}}(q)}\prod_{k=1}^n(z_k-q).
 \end{equation}
We start rewriting $\tau_\mathcal{C}$ with the $x$-dependent part, we use presentation  Eq. \eqref{xi22} along with \eqref{omegaomega2}
\begin{equation}
	\exp\left(x\int\limits_{\mathcal{C}} \frac{dq}{q}  \nu(q)\right)=e^{x\Omega_{\scaleto{>}{3pt}}(0)} = \frac{e^{x\omega_{\scaleto{>}{3pt}}(0)}}{\prod\limits_{k=1}^n z_k^{x} }=\exp\left(x\int\limits_{\mathds{S}^1}\frac{dq}{q}\nu_{\mathfrak{n}} (q)\right)\prod_{k=1}^n z_k^{-x}.
\end{equation}
Proceeding in a similar way we rewrite the remaining double integral in \eqref{tauCanswer2} (for details of the computations see \eqref{F2}). 
Overall the tau function \eqref{tauCanswer} can be rewritten as 
\begin{equation}\label{tauCreduced}
	\tau_{\mathcal{C}}[\nu] = \frac{\prod\limits_{k=1}^n\prod\limits_{j \neq k}^n(z_j - z_k)}{\prod\limits_{k=1}^n z_k^{2n+x} \theta'(z_k)} \exp\left((x+n)\omega_{\scaleto{>}{3pt}}(0)+\int\limits_{\mathds{S}^1} \frac{dq}{2\pi i}\omega'_{\scaleto{<}{3pt}}(q)\omega_{\scaleto{>}{3pt}}(q) -2\sum\limits_{k=1}^n \omega_{\scaleto{<}{3pt}}(z_k)\right),
\end{equation}
or in terms of $\nu_{\mathfrak{n}}$ 
	\begin{equation}
		\tau_{\mathcal{C}}[\nu] =\exp\left(S[\nu_{\mathfrak{n}}]\right) \frac{\prod\limits_{k=1}^n\prod\limits_{j \neq k}^n(z_j - z_k)}{\prod\limits_{k=1}^n z_k^{2n+x} \theta'(z_k)} ,
	\end{equation}
 with \begin{equation}
   S[\nu_{\mathfrak{n}}] = -(x+n)\int\limits_{\mathds{S}^1}\ln(q)d\nu_{\mathfrak{n}} (q)+ \int\limits_{\mathds{S}^1}\int\limits_{\mathds{S}^1}  \ln|k-q|d\nu_{\mathfrak{n}}(k)d\nu_{\mathfrak{n}}(q)+2\sum\limits_{j=1}^n \int\limits_{\mathds{S}^1}  \ln(z_j-k)d\nu_{\mathfrak{n}}(k)
 \end{equation}
	here $d\nu_{\mathfrak{n}}(q) \equiv \nu_{\mathfrak{n}}'(q)dq$. 
	If we wish further to present an answer in terms of  $\nu(q)$ we have to take into account that it is not a single-valued function on 
	$\mathds{S}^1$ and choose a particular integration path. 
	Let us denote by $\hat{\mathds{S}}^1$ the integration contour $q \in \hat{\mathds{S}}^1 \Leftrightarrow q =e^{i\varphi}, \varphi \in [-\pi,\pi)$.
	This choice allows us to conclude that 
		\begin{equation}
		\int\limits_{\hat{\mathds{S}}^1}\frac{dq}{q} \ln q =-      \int\limits_{-\pi}^\pi \varphi d\varphi =0,
	\end{equation}
	which along with  
	\begin{equation}
		\frac{1}{2\pi i} \int\limits_{\hat{\mathds{S}}^1} \frac{dq}{q} \ln(z_k-q)=\ln z_k, \qquad 
		\int\limits_{\hat{\mathds{S}}^1}\frac{dp}{p}\ln|q-p| = 0,
	\end{equation}
	leads to  
	\begin{equation}\label{tauCnu}
\boxed{	\tau_{\mathcal{C}}[\nu] =\exp\left(S[\nu]\right)  \frac{\prod\limits_{k=1}^n\prod\limits_{j \neq k}^n(z_j - z_k)}{\prod\limits_{k=1}^n z_k^{x} \theta'(z_k)}.}
\end{equation}
where now 
\begin{equation}
  S[\nu] = -(x+n)\int\limits_{\hat{\mathds{S}}^1}\ln(q)d\nu (q)+ \int\limits_{\hat{\mathds{S}}^1}\int\limits_{\hat{\mathds{S}}^1}  \ln|k-q|d\nu(k)d\nu(q)+2\sum\limits_{j=1}^n \int\limits_{\hat{\mathds{S}}^1}  \ln(z_j-k)d\nu(k).
\end{equation}
The formula \eqref{tauCnu} coincides with the leading terms of the Hartwig-Fisher asymptotic of the corresponding Toeplitz determinant \cite{Hartwig_1969}. We discuss it in more detail below in Sec. \eqref{Sec:effective}, while the subleading corrections are discussed in Sec. \eqref{SubLead}.

\section{Subleading corrections}
\label{SubLead}

The technique developed in the previous section allows us to prove that Eq. \eqref{tauCanswer} indeed gives the leading asymptotic to the Toeplitz determinant \eqref{toeplitzdef} and find the exact relations between $\tau_C[\nu]$ and $T_x[\theta]$ valid for any $x>0$. 
We start by identically presenting 
\begin{equation}
    T_x[\theta]= \det_{\mathcal{C}}(1+\hat{S}) = \det_{\mathcal{C}}(1+\hat{V}_{\mathcal{C}}-\hat{\Delta}),\qquad \hat{\Delta} \equiv \hat{V}_{\mathcal{C}}-\hat{S}. 
\end{equation}
The kernel for $\hat{\Delta}$ is explicitly given by
\begin{equation}
 \Delta(q,p)  =\frac{\sqrt{\theta(p)}\sqrt{\theta(q)}p^{-x/2}q^{-x/2}}{2\pi i}
    \int\limits_{\mathcal{C}^{\scaleto{<}{3pt}}} \frac{dk}{2\pi i} \frac{k^x \theta(k)}{(1+\theta(k))(k-q)(k-p)}
\end{equation}
or equivalently for  $x>0$ 
\begin{equation}
   \Delta(q,p)   = -\frac{\sqrt{\theta(p)}\sqrt{\theta(q)}p^{-x/2}q^{-x/2}}{2\pi i}
    \int\limits_{\mathcal{C}^{\scaleto{<}{3pt}}} \frac{dk}{2\pi i} \frac{k^x}{(1+\theta(k))(k-q)(k-p)}. 
\end{equation}
Notice that $q$ and $p$ dependence of this expression is the same as in the last term of the variational formula Eq. \eqref{var1}.  
Computation of the full determinant $ T_x[\theta]$ can be reduced to the following procedure 
\begin{equation}
       T_x[\theta]=  \det_{\mathcal{C}}(1+\hat{V}_{\mathcal{C}}-\hat{\Delta}) = \tau_{\mathcal{C}} \det_{\mathcal{C}} (1 - (1-\hat{R})\hat{\Delta}).
\end{equation}
Using the cyclic permutation under the determinant and the relations \eqref{Mdef}, and \eqref{fromL2},  as well as Eq. \eqref{tauCanswer} for $\tau_\mathcal{C}$ we can present the Toeplitz determinant as 
	\begin{equation}\label{toep}
	\boxed{	T_x[\theta]=  \det_{\mathcal{C}}(1-\hat{K})\exp\left(x \int\limits_{\mathcal{C}} \frac{dq}{q}  \nu(q) 
		- \frac{1}{2}\int\limits_{\mathcal{C}} \int\limits_{\mathcal{C}} \left(\frac{\nu(k)-\nu(q)}{k-q}\right)^2 dk \, dq\right).}
\end{equation}
with the integrable kernel $K$
\begin{equation}\label{Kkernel}
    K(k_1,k_2) =  
    \frac{k_1^{x}e^{\Omega_{\scaleto{>}{3pt}}(k_1)+\Omega_{\scaleto{>}{3pt}}(k_2)} }{2\pi i(1+\theta(k_1))}  \frac{b_{\scaleto{>}{3pt}}(k_1)-b_{\scaleto{>}{3pt}}(k_2)}{k_1-k_2}.
\end{equation}
Here we have presented $K$ in a form that has allowed us to transform the contour in the Fredholm determinant $\mathcal{C}^{\scaleto{<}{3pt}} \to \mathcal{C}$.

There are a couple ways of how to transform this kernel even further such that the $x$-dependence is manifest. 
We call them the first and the second Slavnov formula, following Ref. \cite{Slavnov_2010}, and Borodin-Okounkov formula after Refs. \cite{Borodin2000,Bttcher2006}.

\begin{figure}[h]
		\centering
	\includegraphics[width=\linewidth]{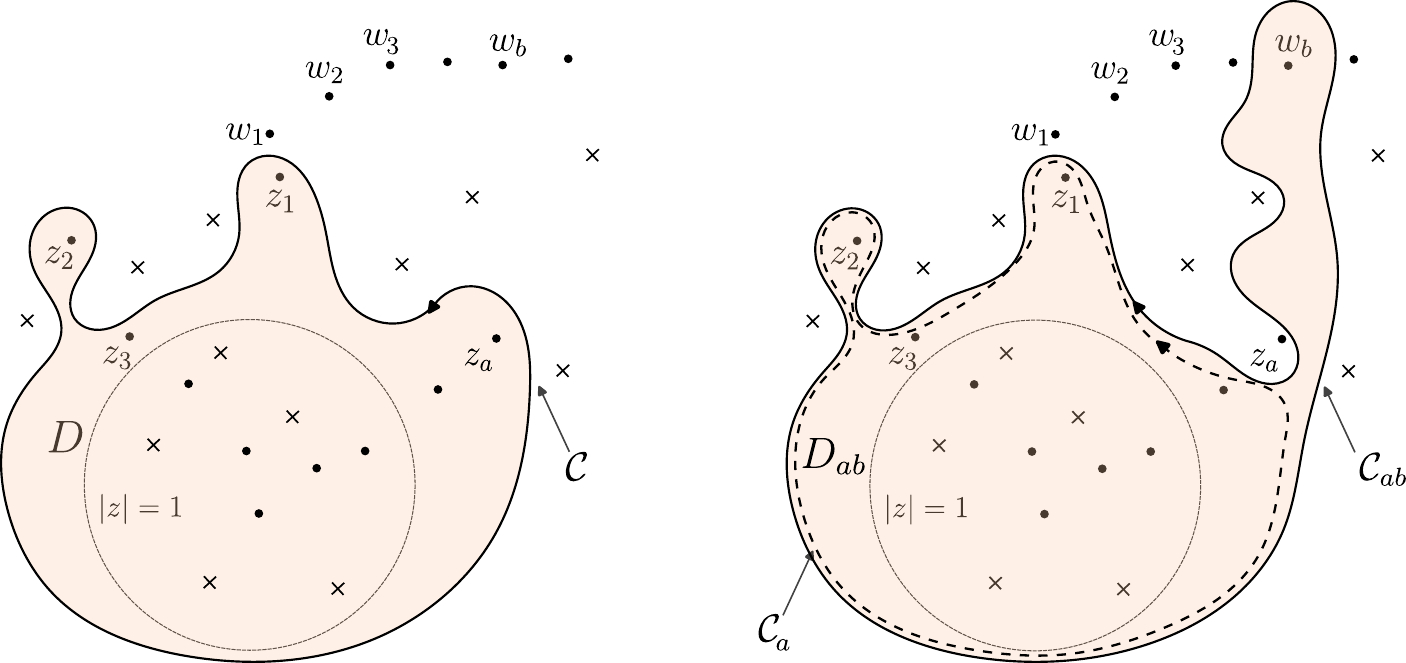}
	\caption{The left panel shows the original contour  $\mathcal{C}$ in the complex $q$ plane, the black dots are zeroes of $1+\theta(q)$ inside the shaded region $D$ - $z_1$, $z_2$, $\dots$, and  outside  - $w_1$, $w_2$, $\dots$. The crosses denote the poles of $\theta(q)$. 
		The right panel shows an example of the contour $\mathcal{C}_{\textbf{z},\textbf{w}}= \mathcal{C}_{z_a,w_b}= \mathcal{C}_{ab}$ obtained by inclusion of the point $w_b$ and exclusion of $z_a$ as well as the intermediate contour $\mathcal{C}_a$ that excludes $z_a$ but does not encircle $w_b$. Notice that in the process of deformation $\mathcal{C} \to \mathcal{C}_a \to\mathcal{C}_{ab}$ neither the poles of $\theta(q)$ are crossed nor other zeroes except $z_a$ and $w_b$. }
	\label{fig2}
\end{figure}

\textit{The first Slavnov formula}. Let us denote solutions of the equation $\theta(q)+1=0$ inside $D$ as $z_n$, $n=1,2,\dots |Z|$ and in $\mathds{CP}^1\setminus D$ as $w_n$, , $n=1,2,\dots |W|$ (see Fig. \eqref{fig2}). 
We allow having an infinite number of zeroes $|W|$ and $|Z|$.
As we have discussed in Sec. \eqref{secS1} all these roots are assumed to be distinct and ordered $\dots >|w_2|>|w_1|>|z_1|> |z_2| >\dots$. 
Then we evaluate $b_{\scaleto{>}{3pt}}(k)$ by shrinking the contour to the infinity  (see \eqref{bpint})
\begin{equation}
    b_{\scaleto{>}{3pt}}(k) =   \int\limits_{\mathcal{C}^{\scaleto{>}{3pt}}} \frac{dq}{2\pi i} \frac{q^{-x}\theta(q) e^{-2\Omega_{\scaleto{<}{3pt}}(q)}}{(1+\theta(q))(k-q)}= \sum\limits_{n=1}^{|W|} \frac{w_n^{-x} e^{-2\Omega_{\scaleto{<}{3pt}}\left(w_n\right)}}{\theta'\left(w_n\right)\left(w_n-k\right)}.
\end{equation}
This results in the following presentation for the kernel \eqref{Kkernel}
\begin{equation}\label{Kkernel1}
     K(k_1,k_2) = \frac{k_1^{x}e^{\Omega_{\scaleto{>}{3pt}}(k_1)+\Omega_{\scaleto{>}{3pt}}(k_2)} }{2\pi i(1+\theta(k_1))}  \sum\limits_{n=1}^{|W|} \frac{\left(w_n\right)^{-x} e^{-2\Omega_{\scaleto{<}{3pt}}\left(w_n\right)}}{\theta'\left(w_n\right)\left(w_n-k_1\right)\left(w_n-k_2\right)}
\end{equation}
Using the cycling property under the determinant we obtain 
\begin{equation}
    \det_{\mathcal{C}}(1-\hat{K}) = \det\limits_{1\le n,m \le |W|}\left(\delta_{nm} - \mathcal{A}_{nm} \right) \equiv  \det\limits_{1\le n,m \le |W|}\left(\delta_{nm} -   \mathcal{A}(w_n,w_m) \right)
\end{equation}
where 
\begin{multline}\label{mathcalA}
   \mathcal{A}(w_n,w_m) = \frac{\left(w_n\right)^{-x} e^{-2\Omega_{\scaleto{<}{3pt}}\left(w_n\right)}}{\theta'\left(w_n\right)} \int\limits_{\mathcal{C}^{\scaleto{<}{3pt}}} \frac{dk}{2\pi i} \frac{\theta(k)e^{2\Omega_{\scaleto{>}{3pt}}(k)}}{1+\theta(k)}\frac{k^x}{\left(w_n-k\right)\left(w_m-k\right)} \\ = - \sum_{l=1}^{|Z|}\frac{\left(w_n\right)^{-x} e^{-2\Omega_{\scaleto{<}{3pt}}\left(w_n\right)}e^{2\Omega_{\scaleto{>}{3pt}}\left(z_l\right)} \left(z_l\right)^x}{\theta'\left(w_n\right)\theta'\left(z_l\right)\left(w_n-z_l\right)\left(w_m-z_l\right)}.
\end{multline}
It can be presented as a matrix product $   \mathcal{A}(w_n,w_m)  = \sum\limits_{l=1}^{|Z|}B(w_n,z_l)C(z_l,w_m)$ with
\begin{equation}
    B(w_n,z_l) = \frac{w_n^{-x} e^{-2\Omega_{\scaleto{<}{3pt}}\left(w_n\right)}} {\theta'\left(w_n\right)\left(w_n-z_l\right)},\qquad    C(z_l,w_m) =
    \frac{e^{2\Omega_{\scaleto{>}{3pt}}\left(z_l\right)} \left(z_l\right)^x}{\theta'\left(z_l\right)\left(z_l-w_m\right)}.
\end{equation}
Notice that under the determinant we can exchange $B\leftrightarrow C$, which is equivalent to exchanging the role of zeroes $w\to z$. Therefore, without loss of generality, we can assume that $|W| \le |Z|$ and use the expansion in traces of the antisymmetric powers of $\mathcal{A}$, namely 
\begin{equation}
    \det\limits_{1\le n,m \le |W|}\left(\delta_{nm} - \mathcal{A}_{nm} \right) = \sum\limits_{k=0}^{|W|} \sum\limits_{{\bf w}, |{\bf w}|=k}\det\limits_{w_n, w_m \in {\bf w}}\left(-\mathcal{A}(w_n,w_m) \right) 
\end{equation}
here ${\bf w} = \{w_{a_1},w_{a_2},\dots w_{a_{k}}\}$ is an ordered set of zeroes $1+\theta(q)$ in $\mathds{CP}^1\setminus D$. Taking into account that $\mathcal{A} = BC$ we can employ the Cauchy-Binet theorem \cite{Shafarevich2013} to present  
\begin{equation}\label{da}
    \det\limits_{1\le n,m \le |W|}\left(\delta_{nm} - \mathcal{A}_{nm} \right) = \sum\limits_{k=0}^{|W|} \sum\limits_{\substack{{\bf w}, {\bf z}\\ |{\bf z}|=|{\bf w}|=k}} \det\limits_{w_n\in {\bf w}, z_l \in {\bf z}}\left(-B(w_n,z_l) \right)\det\limits_{z_l \in {\bf z}, w_m\in {\bf w}}\left(C(z_l,w_m) \right)\equiv\sum\limits_{k=0}^{|W|} \sum\limits_{\substack{{\bf w}, {\bf z}\\ |{\bf z}|=|{\bf w}|=k}} \mathfrak{D}_{{\bf z},{\bf w}}.
\end{equation}
Here  ${\bf z} = \{z_{b_1},z_{b_2},\dots z_{b_{k}}\}$  is an ordered set of zeroes $1+\theta(q)$ in $D$.
The determinants are of the Cauchy type and can be easily evaluated 
\begin{equation}\label{det1}
\mathfrak{D}_{{\bf z},{\bf w}} = 
    \prod\limits_{w_n \in {\bf w}} \frac{w_n^{-x}e^{-2\Omega_{\scaleto{<}{3pt}}(w_n)}}{\theta'(w_n)}\prod\limits_{z_l\in {\bf z}} \frac{z_l^{x}e^{2\Omega_{\scaleto{>}{3pt}}(z_l)}}{\theta'(z_l)} \frac{\prod\limits_{a> b, \, w_a, w_b \in {\bf w}}(w_a-w_b)^2\prod\limits_{a> b, \, z_a, z_b \in {\bf z}}(z_a-z_b)^2}{\prod\limits_{z_l \in {\bf z},\, w_n \in {\bf w}}(z_l-w_n)^2}
\end{equation}
From this expression, it is clear that for $x\to +\infty$ all such determinants vanish, $\det(1-\mathcal{A}) \to 1$. 
This, in particular, is related to the proper choice of the contour $\mathcal{C}$ discussed in Sec. \eqref{secS1}. 

\textit{The second Slavnov formula } relates the subleading terms to the different choices of contour $\mathcal{C}$. Namely, let us denote $\mathcal{C}_{\bf z, w}$ the contour that  (i) is obtained from $\mathcal{C}$ by a smooth deformation avoiding crossing poles of $\theta(q)$; (ii)  excludes  
${\bf z} = \{z_{a_1}, z_{a_2},\dots z_{a_{k}}\}$; and (iii) includes the points ${\bf w} = \{w_{b_1}, w_{b_2},\dots w_{b_{k}}\}$, in the area $D_{\bf{z},\bf{w}}$ encircled by $\mathcal{C}_{\bf{z},\bf{w}}$ (see Fig. \eqref{fig2}). Notice that the winding number of $\mathcal{C}_{\bf{z},\bf{w}}$  is still zero so we can associate the corresponding tau function $\tau_{\mathcal{C}_{\bf z, w}}$ given by \eqref{tauCanswer}. The second Slavnov formula generalized from the real line now reads 
\begin{equation}\label{tauRel}
    \tau_{\mathcal{C}_{\bf z, w}}[\nu]  =      \mathfrak{D}_{{\bf z},{\bf w}}  \tau_{\mathcal{C}}[\nu]. 
\end{equation}
Or equivalently for $T_x[\theta]$ the full asymptotic expansion can be presented as 
\begin{equation}\label{tauCont}
   \boxed{ T_x[\theta] = \sum_{\mathcal{C}_{\bf z, w}} \tau_{{\mathcal{C}_{\bf z, w}} }[\nu]}
\end{equation}
where each $\tau_{\mathcal{C}_{\bf z, w}}[\nu]$ is given by Eq. \eqref{tauCanswer} and summation is over all possible contours with zero winding number of $\nu(q)$ that are obtained from $\mathcal{C}$  by smooth deformations that don't cross poles of $\theta(q)$. 

Let us start proving Eq. \eqref{tauRel}  from the simplest case  $\textbf{z} = \{z_a\}$ and $\textbf{w} = \{w_b\} $ 
i.e. when we deform the contour $\mathcal{C}$ to include/exclude only one point (see Fig. \eqref{fig2}).  
With respect to this new contour $\mathcal{C}_{\bf z, w} \equiv 
\mathcal{C}_{ab}$ we define $\Omega_{\mathcal{C}_{ab}}$ as in Eq. \eqref{bpint}. 
Along the lines of Eqs. \eqref{OmegaPM} we define $	\Omega_{\mathcal{C}_{ab}^{\scaleto{>}{3pt}}}$ to be analytic inside the $\mathcal{C}_{ab}$ (the shaded area $D_{ab}$ in Fig. \eqref{fig2}) and $	\Omega_{\mathcal{C}_{ab}^{\scaleto{<}{3pt}}}$ analytic outside. 
These functions are connected by the same jump as the original scalar RHP  \eqref{scalarRHP} but this time while crossing the  contour $\mathcal{C}_{ab}$
\begin{equation}\label{scalarRHP1}
	\Omega_{\mathcal{C}_{ab}^{\scaleto{>}{3pt}}} (q)-    	\Omega_{\mathcal{C}_{ab}^{\scaleto{<}{3pt}}} (q) = 2\pi i \nu(q). 
\end{equation}
To connect $ \Omega_{\mathcal{C}_{ab}}$ to $\Omega_\mathcal{C}$, we can perform the same manipulations as when deriving \eqref{omegaomega1}, \eqref{omegaomega2} i.e. we compute the derivatives, deform the contour to account for the residues, and then integrate back, accounting for the integration constant from the normalization condition. Overall, we get
\begin{equation}\label{omegaomega3}
	e^{\Omega_{\mathcal{C}_{ab}^{\scaleto{>}{3pt}}}(q)}  = e^{\Omega_{\scaleto{>}{3pt}}(q)}\frac{q-z_a}{q-w_b} \Longrightarrow e^{\Omega_{\mathcal{C}_{ab}^{\scaleto{<}{3pt}}}(q)}  = e^{\Omega_{\scaleto{<}{3pt}}(q)}\frac{q-z_a}{q-w_b}.
\end{equation}
Repeating this procedure we conclude that 
\begin{equation}\label{omegaomega4}
	e^{\Omega_{\mathcal{C}_{{\bf z}, {\bf w}}^{\scaleto{>}{3pt}}}(q)}  = e^{\Omega_{\scaleto{>}{3pt}}(q)} \prod_{z_a\in {\bf z}, {w_b \in {\bf w}}}\frac{q-z_a}{q-w_b} .
\end{equation}
Imagine that we have managed to demonstrate that 
\begin{equation}\label{tauAB1}
	\tau_{\mathcal{C}_{ab}}[\nu] = \frac{z_a^x w_b^{-x} e^{2\Omega_{\scaleto{>}{3pt}}(z_a)-2\Omega_{\scaleto{<}{3pt}}(w_b)}}{\theta'(z_a)\theta'(w_b)(z_a-w_b)^2} \tau_{\mathcal{C}}[\nu].
\end{equation}
Then this relation can be extended to the more general contours $\mathcal{C} \to \mathcal{C}_{\bf z, w}$ and $\mathcal{C}_{ab} \to \mathcal{C}_{{\bf z}\cup z_a,{\bf w}\cup w_b} $. Here we have added  point $z_a$ to ${\bf z}$ and $w_b$ to ${\bf w}$. The functions $\Omega$ should be modified in accordance with \eqref{omegaomega4}. 
Then \eqref{tauAB1} is equivalent to \eqref{det1}.  
Indeed, we can rewrite \eqref{det1} as 
\begin{equation}
\mathfrak{D}_{{\bf z}\cup z_a,{\bf w}\cup w_b} =\frac{z_a^xw_b^{-x}}{\theta'(z_a)\theta'(w_b)(z_a-w_b)^2} \left(
e^{2\Omega_{\scaleto{>}{3pt}}(z_a)} \frac{\prod\limits_{z_k \in {\bf z}}(z_a-z_k)^2}{\prod\limits_{w_k \in {\bf w}}(z_a-w_k)^2}
\right)
\left(
e^{-2\Omega_{\scaleto{<}{3pt}}(w_b)} \frac{\prod\limits_{w_k \in {\bf w}}(w_b-w_k)^2}{\prod\limits_{z_k \in {\bf z}}(w_b-z_k)^2}
\right)
\mathfrak{D}_{{\bf z},{\bf w}}.
\end{equation}
Taking into account \eqref{omegaomega4} the expressions in brackets are nothing but the functions $\Omega$ on the corresponding contours and the whole expression is nothing but \eqref{tauAB1} after the multiplication of both sides by $\tau_{\mathcal{C}}$.

To prove \eqref{tauAB1} we have to can either use the explicit formula \eqref{tauCanswer2} and follow a tedious path of deriving Eq. \eqref{tauCreduced} outlined in  \ref{F2} or obtain it directly from the definition of the tau function  \eqref{tauC}. 
To do the latter, we introduce the contour $\mathcal{C}_a$, which serves as an intermediate between $\mathcal{C}$ and $\mathcal{C}_{ab}$ i.e. the point $z_a$ is already excluded but $w_b$ is not included yet (see Fig. \eqref{fig2}). We can define the corresponding tau function $\tau_{\mathcal{C}_{a}}[\nu] =\det\limits_{\mathcal{C}_{a}}(1 + \hat{V}_{\mathcal{C}_{a}})$, however it is no longer given by Eq. \eqref{tauCanswer2} as the winding of $\nu(q)$ on $\mathcal{C}_a$ is no longer zero.  Nevertheless, from the definition of the kernel \eqref{Vdef1} and \eqref{Vdef2} we see that the contour $\mathcal{C}_a$ can be extended to both $\mathcal{C}$ and $\mathcal{C}_{ab}$ because it lies inside each of them so moving $q$ in $w_{\mathcal{C}_a}$ in Eq. \eqref{Vdef2} would never cross the integration contour. 
This way  
\begin{equation}
	\tau_{\mathcal{C}_{a}}[\nu] =\det\limits_{\mathcal{C}_{a}}(1 + \hat{V}_{\mathcal{C}_{a}})=\det\limits_{\mathcal{C}_{ab}}(1 + \hat{V}_{\mathcal{C}_{a}})= \det\limits_{\mathcal{C}_{ab}} (1 + \hat{V}_{\mathcal{C}_{ab}}+\hat{W}_{w_b})  ,
\end{equation}
\begin{equation}
	\tau_{\mathcal{C}_{a}}[\nu] =\det\limits_{\mathcal{C}_{a}}(1 + \hat{V}_{\mathcal{C}_{a}})=\det\limits_{\mathcal{C}}(1 + \hat{V}_{\mathcal{C}_{a}}) = \det\limits_{\mathcal{C}} (1 + \hat{V}_{\mathcal{C}}+\hat{W}_{z_a}) ,
\end{equation}
where the additional terms come from the difference between $w_{\mathcal{C}_{a}}$ and $w_{\mathcal{C}_{ab}} $ with  $w_\mathcal{C}$ in \eqref{Vdef2}. They can be computed as residues in the corresponding points
\begin{equation}
	W_{s}(q,p) = \frac{\sqrt{\theta(q)\theta(p)}}{2\pi i}  \frac{s^xq^{-x/2} p^{-x/2} }{\theta'(s)(s-q)(s-p)}.
\end{equation}
On both contours, the winding number is zero so we can use the exact formulas for the resolvent along with Eq. \eqref{Mdef} to evaluate the contributions of the rank-1 corrections explicitly. 
We have 
\begin{equation}
		\frac{\tau_{\mathcal{C}_{a}}[\nu]}{\tau_\mathcal{C}[\nu]} = 1 + {\rm Tr} (1-\hat{R})\hat{W}_{z_a} =  1 + \frac{z_a^x}{\theta'(z_a)} M(z_a,z_a) \overset{\eqref{fromL2}, \eqref{bpint}}{=}  1+ \int\limits_{\mathcal{C}} \frac{ds}{2\pi i} \left(\frac{s}{z_a}\right)^{-x} \frac{\theta(s) e^{2\Omega_{\scaleto{>}{3pt}}(z_a)-2\Omega_{\scaleto{<}{3pt}}(s)}}{(1+\theta(s))\theta'(z_a)(s-z_a)^2}.
\end{equation}
Using the scalar RHP \eqref{scalarRHP} we present an integrand in a form that allows one to deform the integration contour back to $\mathcal{C}_a$. By doing so we have to take into account the residue at $s=z_a$, namely, 
\begin{equation}\label{ta}
		\frac{\tau_{\mathcal{C}_{a}}[\nu]}{\tau_\mathcal{C}[\nu]} = \frac{z_a^xe^{2\Omega_{\scaleto{>}{3pt}}(z_a)}}{\theta'(z_a)}\int\limits_{\mathcal{C}_a} \frac{ds}{2\pi i}s^{-x} \frac{(1+\theta(s))\theta(s) e^{-2\Omega_{\scaleto{>}{3pt}}(s)}}{(s-z_a)^2} 
\end{equation}
And in a similar way   
\begin{equation}\label{tab}
	\frac{\tau_{\mathcal{C}_{a}}[\nu]}{\tau_{\mathcal{C}_{ab}}[\nu]} =    \frac{w_b^xe^{2\Omega_{\mathcal{C}^{\scaleto{>}{3pt}}_{ab}}(w_b)}}{\theta'(w_b)}\int\limits_{\mathcal{C}_a} \frac{ds}{2\pi i} s^{-x} \frac{(1+\theta(s))\theta(s) e^{-2\Omega_{\mathcal{C}^{\scaleto{>}{3pt}}_{ab}}(s)}}{(s-w_b)^2} .
\end{equation}
Taking into account Eq. \eqref{omegaomega3} we notice that the integrals in both Eqs. \eqref{ta} and \eqref{tab} are the same, which leads to 
\begin{equation}\label{zh1}
	\frac{\tau_{\mathcal{C}_{ab}}[\nu]}{\tau_\mathcal{C}[\nu]} = \frac{z_a^x w_b^{-x} \theta'(w_b)e^{2\Omega_{\scaleto{>}{3pt}}(z_a)}  }{\theta'(z_a)e^{2\Omega_{\mathcal{C}^{\scaleto{>}{3pt}}_{ab}}(w_b)} }.
\end{equation}
The value in the denominator should be understood as a limit. Namely, from \eqref{omegaomega3} we get 
\begin{equation}
		e^{\Omega_{\mathcal{C}_{ab}^{\scaleto{>}{3pt}}}(q)}  =(1+\theta(q)) \frac{q-z_a}{q-w_b} e^{\Omega_{\scaleto{<}{3pt}}(q)}
\end{equation}
Thus for $q=w_b$ we get 
\begin{equation}
	e^{\Omega_{\mathcal{C}_{ab}^{\scaleto{>}{3pt}}}(w_b)}  =\lim\limits_{q\to w_b}(1+\theta(q)) \frac{q-z_a}{q-w_b} e^{\Omega_{\scaleto{<}{3pt}}(q)} = (w_b-z_a)\theta'(w_b)e^{\Omega_{\scaleto{<}{3pt}}(w_b)}.
\end{equation}
Substituting this expression to \eqref{zh1} we recover \eqref{tauAB1}. 

{\it Borodin-Okounkov formula}. Let us briefly comment on how to present the Fredholm determinant $\det_\mathcal{C}(1-\hat{K})$ acting on the contour as a Fredholm determinant acting on $ l_2({x,x+1,...})$. 
We focus only on the case $\mathcal{C} = \mathds{S}^1$. 
First we note that functions $\Omega_{\scaleto{>}{3pt}}$ and $\Omega_{\scaleto{<}{3pt}}$ \eqref{OmegaPM} solve the factorization problem for the symbol $\phi(q) = 1 + \theta(q)$ on the contour $\mathcal{C}$, namely
\begin{equation}
    \phi_+(q)\phi_-(q)=\phi(q)=1+\theta(q)=e^{\Omega_{\scaleto{>}{3pt}}(q)}e^{-\Omega_{\scaleto{<}{3pt}}(q)},
\end{equation}
with the identification
\begin{equation}
  \phi_+(q) = e^{\Omega_{\scaleto{>}{3pt}}(q)},\qquad   \phi_-(q)=e^{-\Omega_{\scaleto{<}{3pt}}(q)}.
\end{equation}
Then the kernel $\hat{K}$ in \eqref{Kkernel} after the conjugation with the diagonal matrices and the slight shift of the contour can be presented as 
\begin{equation}
    K(k_1,k_2)=\frac{1}{2\pi i}\int\limits_{\mathcal{C}^{\scaleto{>}{3pt}}}\frac{ds}{2\pi i} A(k_1,s)B(s,k_2),
\end{equation}
with 
\begin{equation}
   A(k_1,s)\equiv \frac{k_1^{x}\phi_+(k_1)\phi_-^{-1}(k_1)}{s-k_1}, \qquad B(s,k_2)\equiv \frac{s^{-x}\phi_+^{-1}(s)\phi_-(s)}{s-k_2}.
\end{equation}
We expand these functions in the Laurent series 
\begin{equation}
    A(k_1,s)=\sum_{n=1}^{\infty} A_n(k_1) s^{-n}, \qquad B(s,k_2)=\sum_{m=-\infty}^{\infty} B_m(k_2)s^{-m-1},
\end{equation}
to present the kernel in terms of $A_n, B_m$, namely 
\begin{equation}
    K(k_1,k_2)=\frac{1}{2\pi i}\sum_{n=1}^{\infty}A_n(k_1)B_{-n}(k_2).
\end{equation}
Using the cyclic property of determinant we can get
\begin{equation}
    \det_{\mathcal{C}}(1-K)=\det_{\mathds{N}}(1-\mathcal{K}),
\end{equation}
where
\begin{equation}
\mathcal{K}_{nm}=\int\limits_{\mathcal{C}}\frac{dk}{2\pi i}B_{-n}(k)A_m(k).
\end{equation}
The explicit form of the Laurent coefficient reads 
\begin{equation}
    A_m(k)=k^{x+m-1}\phi_+(k)\phi_-^{-1}(k),
\end{equation}
\begin{equation}
    B_{-n}(k)=\int\limits_{\mathcal{C}^{\scaleto{>}{3pt}}}\frac{s^{-n}ds}{2\pi i}\frac{s^{-x}\phi_+^{-1}(s)\phi_-(s)}{s-k}=\sum_{l=0}^{\infty}k^l (\phi_+^{-1}\phi_-)_{x+n+l}.
\end{equation}
Therefore 
\begin{equation}
\mathcal{K}_{nm}=\sum_{l=0}^{\infty}(\phi_+^{-1}\phi_-)_{x+n+l}\int\limits_{\mathcal{C}}\frac{dk}{2\pi i}k^{x+m+l-1}\phi_+(k)\phi_-^{-1}(k)=\sum_{l=0}^{\infty}(\phi_+^{-1}\phi_-)_{x+n+l}(\phi_+\phi_-^{-1})_{-x-m-l}.
\end{equation}
which is after the shift of the $n$ and $m$ turns into a kernel acting on $ l_2({x,x+1,...})$. 

\section{Effective form factors}
\label{Sec:effective}

Finally, let us comment on the heuristic approach of the effective form factors. In \cite{10.21468/SciPostPhys.10.3.070} we have considered a formal form factor series 
\begin{equation}\label{tauEffSeries}
    \tau^{\rm eff}[\nu] \equiv \sum_{\mathbf{q}} |\langle {\bf p}|{\bf q}\rangle|^2 \prod\limits_{i=1}^N \frac{q_i^x}{p_i^x},
\end{equation}
where the ordered set $\textbf{p}=\{p_1,\dots p_{N}\}$ of the \textit{shifted} momenta consists of $N$ distinct solutions of the equation $p^L = e^{-2\pi i \nu(p)}$; the set $\mathbf{q} = \{q_1,\dots, q_{N}\}$ is taken from the solutions of the \textit{unshifted} equation $q^L=1$; and the form factors are declared to be 
\begin{equation}\label{pq}
    |\langle {\bf p}|{\bf q}\rangle|^2  =\left(\frac{2}{L}\right)^{2N}\prod\limits_{i=1}^N \frac{p_i q_i \theta(q_i)\theta(p_i)}{(1+\theta(q_i))(1+\frac{2\pi}{L}\nu'(p_i))} 
	\frac{\prod\limits_{i>j}^{N}(p_i-p_j)^2\prod\limits_{i>j}^{N}(q_i-q_j)^2}{\prod\limits_{i,j=1}^{N}(p_i-q_j)^2}.
\end{equation}
Here we have changed notations to exponential ones $e^{ip}\to p$ compared to \cite{10.21468/SciPostPhys.10.3.070}. The main result of  \cite{10.21468/SciPostPhys.10.3.070} is that in the thermodynamic limit 
$N\to\infty$, $N/L\to 1$, on the one hand the sum in \eqref{tauEffSeries} converges into to the Fredholm determinant 
\begin{equation}\label{tauEffDet}
    \tau^{\rm eff}[\nu] = \det\limits_{\mathds{S}^1}\left(
     1 + \hat{V}_{\mathds{S}^1}
    \right)
\end{equation}
with the kernels defined in \eqref{Vdef1}, \eqref{Vdef2}. On the other hand, we can apply the thermodynamic limit to each element of the sum \eqref{tauEffSeries} and evaluate it afterward. 
This leads to elementary expressions. For example, it is clear that if the winding number is zero we can put $N=L$ so that we end up with only one term in the sum \eqref{tauEffSeries}. More specifically, let us consider non-positive winding number $\mathfrak{n} = -n \le 0$, then (Eqs. (46),(105) in \cite{10.21468/SciPostPhys.10.3.070})
\begin{equation}\label{tauHF}
    \tau^{\rm eff}[\nu] =\det_{1\leq i,j \leq n} y_{\mathfrak{n}}(x+i-j)\times\exp\left((x+n)\int\limits_{\mathds{S}^1}\frac{dq}{q}\nu_{\mathfrak{n}} (q)-\frac{1}{2}\int\limits_{\mathds{S}^1}\int\limits_{\mathds{S}^1} dqdp\left(\frac{\nu_{\mathfrak{n}}(q)-\nu_{\mathfrak{n}}(p)}{q-p}\right)^2\right),
\end{equation}
here 
\begin{equation}
	y_{\mathfrak{n}}(s)=\frac{1}{2\pi i}\int\limits_{\mathds{S}^1} \frac{dk}{k} k^{-s} e^{-2\pi i\nu_{\mathfrak{n}}(k)}e^{-2\omega_{\scaleto{<}{3pt}}(k)}.
\end{equation}
This formula is nothing but the Hartwig-Fisher asymptotic (Theorem~4 in Ref. \cite{Hartwig_1969}) for the  Toeplitz determinant with the symbol related to  
$\nu(q)$ as in Sec. \eqref{secIntro}). However, let us emphasize again that Eq. \eqref{tauHF} is an exact result for the sum \eqref{tauEffSeries} in the thermodynamic limit (i.e. for the determinant \eqref{tauEffDet}) valid for any $x>0$ (not just asymptotically). 
Moreover, in  \eqref{tauHF:app} we prove {Eq.~\eqref{tauHF}} without referring to the form factor series but directly from the Riemann-Hilbert approach.
Presenting the asymptotic expansion for each of $y_{\mathfrak{n}}(x)$ and computing the leading asymptotic in \eqref{tauHF}  one can recover \eqref{tauCnu}. 
It is instructive, however, to discuss the relationship between $\tau^{\rm eff} $ and $\tau_{\mathcal{C}}  $ directly from their definitions \eqref{tauC} and \eqref{tauEffDet} correspondingly. First notice that the definition of the kernel  \eqref{Vdef1} allows us to deform the contour in $\tau^{\rm eff}$ from $\mathds{S}^1$ to $\mathcal{C}$. Indeed, by definition, during the deformation process, we never cross poles of $\theta(q)$, and integration of $w_{\mathds{S}^1}$ in   \eqref{Vdef2} can be continued for $q$ outside the unit circle, similarly as we did in Sec. \eqref{SubLead}.

On the other hand,
\begin{equation}\label{deltaV}
    \hat{V}_{\mathds{S}^1} = \hat{V}_{\mathcal{C}} + \hat{W},\qquad 
    W(p,q) =  \frac{\sqrt{\theta(q)\theta(p)}q^{-x/2}p^{-x/2}}{2\pi i} \sum\limits_{j=1}^n\frac{z_j^{x}}{\theta'(z_j)(z_j-q)(z_j-p)} 
\end{equation}
On the contour $\mathcal{C}$ we can invert $1+\hat{V}_\mathcal{C}$ to obtain 
\begin{equation}\label{taueff}
    \tau^{\rm eff}[\nu] =  \det\limits_{\mathcal{C}}\left(
     1 + \hat{V}_{\mathds{S}^1}
    \right) =  \det\limits_{\mathcal{C}}\left(
     1 + \hat{V}_{\mathcal{C}} + \hat{W} 
    \right) = \tau_{\mathcal{C}}[\nu] \det\limits_{\mathcal{C}}\left(1+ (1-\hat{R})\hat{W}\right) 
\end{equation}
Further, we can use the cyclic property of determinant along with Eqs. \eqref{Mdef} and \eqref{fromL2} to present the Fredholm determinant  
$\det\left(1+ (1-\hat{R})\hat{W}\right) $ as the determinant with the kernel of $n\times n$ matrix $\det_{1\le i,j \le n}(\delta_{ij}-K(z_i,z_j))$ with 
the kernel defined as ({\it cf} \eqref{Kkernel} and \eqref{Kkernel1}) 
\begin{equation}
      K(z_i,z_j) =-\frac{z_j^x}{\theta'(z_j)}M(z_i,z_j)=\frac{z_j^x}{\theta'(z_j)} e^{\Omega_{\mathcal{C}}(z_i)+\Omega_{\mathcal{C}}(z_j)} \frac{b_{\mathcal{C}}(z_i)-b_{\mathcal{C}}(z_j)}{z_i-z_j}.
\end{equation}
To compare with the results of the Sec. \eqref{SubLead} we can do a cyclic transformation once again to exchange $z\leftrightarrow w$, which leads to
\begin{equation}
    \tau^{\rm eff}[\nu] =  \tau_{\mathcal{C}}[\nu]\det\limits_{1\le n,m \le |W|}\left(\delta_{nm} -   \widetilde{\mathcal{A}}(w_n,w_m) \right)
\end{equation}
with 
\begin{equation}
   \widetilde{\mathcal{A}}(w_n,w_m) =  - \sum_{l=1}^{|\delta|}\frac{w_n^{-x} e^{-2\Omega_{\scaleto{<}{3pt}}\left(w_n\right)}e^{2\Omega_{\scaleto{>}{3pt}}\left(z_l\right)} z_l^x}{\theta'\left(w_n\right)\theta'\left(z_l\right)\left(w_n-z_l\right)\left(w_m-z_l\right)}
\end{equation}
So contrary to Eq. \eqref{mathcalA} where the summation is over all the solutions inside the area $D$ encircled by the contour $\mathcal{C}$, 
here the summation is only over $z_k$ which lie between the contour $\mathcal{C}$ and the unit circle (the dashed circle labeled by $"|z|=1"$ in Fig. \eqref{fig2}). 
Similarly, we can write the analog of the second Slavnov formula \eqref{tauCont}. The overall form remains the same $\tau^{\rm eff}[\nu] = \sum_{\mathcal{C}_{\bf z, w}} \tau_{{\mathcal{C}_{\bf z, w}} }[\nu]$, however the summation is limited to the sets  $\mathbf{z}$ consisting only on the points between $\mathcal{C}$ and the unit circle. 
This way, the effective form factors give the leading asymptotic up to the order $O(e^{-x(|w_{|\delta|+1}|+|z_{|\delta|+1}|)})$. 

Throughout the paper, we focused on the negative winding numbers. Still, the final answers \eqref{tauCanswer}, \eqref{tauCanswer2} remain unchanged up to the correct choice of the contour $\mathcal{C}$ for any winding number. For the effective form factor approach the situation is a bit different. As we suggest in  \cite{10.21468/SciPostPhys.10.3.070} the effective form factor approach for the positive winding of $\nu(q)$ can be achieved by a slight change in the definition of $\tau_{\mathcal{C}}$, namely
\begin{equation}
    \tilde{\tau}^{\rm eff}[\nu] = \det\limits_{\mathds{S}^1}\left(
     1 + \hat{Q}_{\mathds{S}^1}
    \right)
\end{equation}
where $\hat{Q}$ is an integral operator acting on $\mathds{S}^1$ with the following kernel
\begin{equation}
	Q(q,p) =\frac{\sqrt{\theta(p)}\sqrt{\theta(q)}}{2\pi i}
	\frac{w_+(p)w_-(q)-w_+(q)w_-(p)}{p-q},
\end{equation}
and function $w_\pm$ are defined as
\begin{equation}
	w_+(q)=q^{x/2}, \qquad w_-(q)=w_+(q)\tilde{w}_{\mathds{S}^1} (q), \qquad 
	\tilde{w}_{\mathds{S}^1} (q) = q^{-x}- \int\limits_{|k|=1+0}\frac{dk}{2\pi i} \frac{k^{-x} \theta(k)}{(1+\theta(k))(k-q)}.
\end{equation}


Another result of \cite{10.21468/SciPostPhys.10.3.070} that readily follows from the \textit{microscopic} presentation  \eqref{tauEffSeries} is the determinants with the special rank-1 correction relevant for the correlation functions. Namely, 
\begin{equation}\label{sd0}
    \det\limits_{\mathds{S}^1}(1+ \hat{V}_\nu +  \hat{V}^{(1)}_\nu ) - 
        \det\limits_{\mathds{S}^1}(1+ \hat{V}_\nu) = \det\limits_{\mathds{S}^1}(1+ \hat{V}_{\nu_1})
\end{equation}
where $\nu_1$ is obtained from $\nu$ by \eqref{nudeltadef} and  $V^{(1)}(p,q) = \sqrt{\theta(p)\theta(q)}p^{-x/2}q^{-x/2}/(2\pi i p)$.
It is interesting that the same relation holds even for the full sine kernel 
\begin{equation}\label{sd}
    \det\limits_{\mathds{S}^1}(1+ \hat{S}_\nu +  \hat{V}^{(1)}_\nu ) - 
        \det\limits_{\mathds{S}^1}(1+ \hat{S}_\nu) = \det\limits_{\mathds{S}^1}(1+ \hat{S}_{\nu_1})
\end{equation}
where the kernel $\hat{S}$ is defined in Eq. \eqref{toeplitzdef}, with the subscript indicating the corresponding $\theta$ via the relation \eqref{nuTheta}. Eq. \eqref{sd} can be demonstrated by rewriting the left and the right parts of the equations as Toeplitz determinants (see Sec. 5 in \cite{10.21468/SciPostPhys.10.3.070}).  Demonstrating the validity of \eqref{sd0} is relatively simple from the microscopic point of view \eqref{tauEffSeries} but quite challenging in the form of Fredholm determinants. We illustrate it below for the particular case of a zero winding number of $\nu$ on $\mathds{S}^1$. In this case, Eq. \eqref{tauHF} gives 
\begin{equation}
     \det\limits_{\mathds{S}^1}(1+ \hat{V}_\nu)  =\exp\left(x\int\limits_{\mathds{S}^1}\frac{dq}{q}\nu (q)-\frac{1}{2}\int\limits_{\mathds{S}^1}\int\limits_{\mathds{S}^1} dqdp\left(\frac{\nu(q)-\nu(p)}{q-p}\right)^2\right)
\end{equation}
and 
\begin{equation}
   \det\limits_{\mathds{S}^1}(1+ \hat{V}_{\nu_1})  =\frac{1}{2\pi i}\int\limits_{\mathds{S}^1} \frac{dk}{k^{x+1}} \frac{e^{-2\Omega_{\scaleto{<}{3pt}}(k)}}{1+\theta(k)}  \exp\left((x+1)\int\limits_{\mathds{S}^1}\frac{dq}{q}\nu (q)-\frac{1}{2}\int\limits_{\mathds{S}^1}\int\limits_{\mathds{S}^1} dqdp\left(\frac{\nu(q)-\nu(p)}{q-p}\right)^2\right),
\end{equation}
or in other words 
\begin{equation}\label{fr}
   \det\limits_{\mathds{S}^1}(1+ \hat{V}_{\nu_1})  =\frac{    \det\limits_{\mathds{S}^1}(1+ \hat{V}_\nu) }{2\pi i}\int\limits_{\mathds{S}^1} \frac{dk}{k^{x+1}} \frac{e^{\Omega_{\scaleto{>}{3pt}}(0)-2\Omega_{\scaleto{<}{3pt}}(k)}}{1+\theta(k)}   =  \det\limits_{\mathds{S}^1}(1+ \hat{V}_\nu)e^{\Omega_{\scaleto{>}{3pt}}(0)} b_{\scaleto{>}{3pt}}(0).
\end{equation}
Here we have used that $[\nu_1]_{-1} = \nu$. 
The resolvent is explicitly given for $1+V_{\nu}$ by Eqs. (\ref{r0}, \ref{fmfp}). 
Moreover, to be able to immediately use formulas (\ref{Mdef}, \ref{fromL2}) we present 
\begin{equation}
     V^{(1)}_\nu(p,q) =\lim\limits_{z_\infty\to \infty} (-z_{\infty})\frac{\sqrt{\theta(p)\theta(q)}}{2\pi i} \frac{p^{-x/2}q^{-x/2}}{p(p-z_{\infty})} . 
\end{equation}
Using the fact that the l.h.s of Eq. \eqref{sd0} is \textit{linear} in $V^{(1)}_\nu$, we obtain 
\begin{equation}
        \det\limits_{\mathds{S}^1}(1+ \hat{V}_\nu + \hat{V}^{(1)}_\nu) - 
        \det\limits_{\mathds{S}^1}(1+ \hat{V}_\nu)  =  \det\limits_{\mathds{S}^1}(1+ \hat{V}_\nu)  
        \lim\limits_{z_\infty\to \infty} (-z_{\infty}) M(0,z_\infty) = \det\limits_{\mathds{S}^1}(1+ \hat{V}_\nu)  e^{\Omega_{\scaleto{>}{3pt}}(0)} b_{\scaleto{>}{3pt}}(0),
\end{equation}
which is exactly as \eqref{fr}. It would be nice to prove \eqref{sd0} directly in the full generality. 

\section{Summary and outlook}

To summarize, we have found closed expressions for the large $x$ asymptotic of the finite temperature sine kernel Fredholm determinants. 
To address the non-zero winding number we have considered the deformation of the contours on which the original determinant is defined along with the deformation of the kernel that corresponds to the effective form factors approach. To clarify this connection was our original motivation. 
In \eqref{tauHF:app} we have presented an alternative way, where one keeps the contour, deforms only the kernel, and still gets a completely solvable Riemann-Hilbert problem. In the end, our approach turned out to be similar to the approach in \cite{Slavnov_2010}, but the advantage of our case is that we have found the resolvent explicitly. Notice that we could have considered more general kernels by deforming $v_\pm(q) \to v_{\pm}(q) e^{\pm g(q)}$. 
Even though the connection with the Toeplitz determinant would be lost in this case, all the procedures for finding the resolvent and asymptotic remain mostly unchanged. However, achieving the full asymptotic expansion might be unreachable because of the properties of the function $g(\lambda)$, which need to be specified case by case. 

The variational formula was an important step in our derivation. This, in our opinion, is the best approach to deal with the generic function $\nu(q)$, 
while in specific cases the differentiation over particular parameters might be preferable. 
Together with differentiation over $x$ this might lead to the closed system of differential equations \cite{Its1990,Korepin1993,Claeys2024,Dean,Dean2019,Cafasso2021,Liechty2020}.

It would be interesting to generalize our approach to the special cases of the so-called Toeplitz + Hankel determinants \cite{Its2020}, 
which seems to be possible since the analogs of the Borodin-Okounkov formula exist in these cases \cite{Betea2018}. 
Another interesting problem is how to add time dependence to the picture. The main obstacle for implementing this also originates from the 
effective form factors and lies in the problems of choosing the continuous effective phase-shift $\nu(q)$ \cite{10.21468/SciPostPhysCore.5.1.006,Zhuravlev2022}. 

{\bf Acknowledgements}. We are grateful to N. Iorgov for many insightful comments. We thank A. Kosyak for useful suggestions on how to improve the presentation.

\appendix 

\section{Lemma about the integrals with the resolvent} \label{lemmaAPP}

In this appendix, we prove Eq. \eqref{fromL2}. To do this we use the explicit form of $f_\pm$ \eqref{fmfp}. 
We start by introducing $r(q;k)$
 \begin{equation}\label{r_def}
  \sqrt{\theta(q)} r(q;k) \equiv    \int\limits_{\mathcal{C}} dp (1-R)(q,p) \frac{\sqrt{\theta(p)}p^{-x/2}}{k-p}  .
 \end{equation}
This quantity defines $M(k_1,k_2)$ in Eq. \eqref{fromL2}  as 
\begin{equation}\label{fromL0}
	M(k_1,k_2)\equiv \int\limits_{\mathcal{C}} \frac{dq}{2\pi i} \frac{q^{-x/2}\theta(q) r(q;k_1)}{k_2 -q} .
\end{equation}
Let us demonstrate how $r(q;k)$ can be expressed via the solution of the RHP \eqref{rhsol}. 
The definition \eqref{r_def} is equivalent to 
\begin{equation}
 r(q;k) \equiv    \frac{ \text{v}_-(q)}{k-q} - \int\limits_{\mathcal{C}} \frac{\theta(p) dp}{2\pi i } \frac{f_+(p)f_-(q) - f_+(q)f_-(p)}{p-q} \frac{ \text{v}_-(p)}{k-p}.
\end{equation}
Then using 
\begin{equation}\label{trick}
    \frac{1}{p-q} \frac{1}{k-p} =  \frac{1}{p-q} \frac{1}{k-q}+ \frac{1}{k-p} \frac{1}{k-q}, 
\end{equation}
we transform $r(q,k)$ as 
\begin{multline}
 r(q;k) \equiv  \frac{1}{k-q}\left(  \text{v}_-(q) - \int\limits_{\mathcal{C}} \frac{\theta(p) dp}{2\pi i } \frac{f_+(p)f_-(q) - f_+(q)f_-(p)}{p-q} \text{v}_-(p) \right)\\
 -\frac{f_-(q)}{k-q}\int\limits_{\mathcal{C}} \frac{\theta(p) dp}{2\pi i }  \frac{ \text{v}_-(p)f_+(p)}{k-p}+
 \frac{f_+(q)}{k-q}\int\limits_{\mathcal{C}} \frac{\theta(p) dp}{2\pi i } \frac{f_-(p) \text{v}_-(p)}{k-p}
\end{multline}
Using Eq. \eqref{int} for the first group of terms and the definition for the RHP \eqref{chiint} for the last two we obtain 
\begin{equation}\label{rqk}
    r(q;k) = \frac{f_-(q)\chi_{22}(k)+f_+(q)\chi_{12}(k)}{k-q}=  \frac{e^{\Omega_{\mathcal{C}}(k)}(f_-(q) + f_+(q) b_{\mathcal{C}}(k))}{k-q},
\end{equation}
with $f_\pm$ given in \eqref{fmfp} and $b_\mathcal{C}$. 

Now let us come back to the integral \eqref{fromL0}. Using the same trick as above \eqref{trick} we obtain (recall that $\text{v}_-(q) = q^{-x/2}$)
\begin{multline}\label{fromL1}
       M(k_1,k_2) = \frac{\chi_{22}(k_1)\chi_{12}(k_2)+\chi_{12}(k_1)(1-\chi_{22}(k_2))-\chi_{22}(k_1)\chi_{12}(k_1)-\chi_{12}(k_1)(1-\chi_{22}(k_1))}{k_1-k_2} \\ = 
       \frac{\chi_{22}(k_1)\chi_{12}(k_2)-\chi_{12}(k_1)\chi_{22}(k_2)}{k_1-k_2} =  - e^{\Omega_{\mathcal{C}}(k_1)+\Omega_{\mathcal{C}}(k_2)} \frac{b_{\mathcal{C}}(k_1)-b_{\mathcal{C}}(k_2)}{k_1-k_2}.
\end{multline}
Here at the last step, we have employed the exact solution \eqref{rhsol}.

\section{Double integral transformation}
\label{F2}

To transform the double integral in \eqref{tauCanswer2} we first identically split it in two parts using Eq. \eqref{omegaomega1} 
\begin{equation}
	\int\limits_{\mathcal{C}} \frac{dq}{2\pi i}\Omega'_{\scaleto{<}{3pt}}(q)\Omega_{\scaleto{>}{3pt}}(q) = J_1 + J_2,
\end{equation}
with 
\begin{equation}
J_1 =-\int\limits_{\mathcal{C}} \frac{dq}{2\pi i}\omega_{\scaleto{<}{3pt}}(q)\Omega'_{\scaleto{>}{3pt}}(q),\qquad J_2= \int\limits_{\mathcal{C}} \frac{dq}{2\pi i}\left(\frac{n}{q}
	-\sum_{k=1}^n \frac{1}{q-z_k}\right)\Omega_{\scaleto{>}{3pt}}(q).
\end{equation}
Here in $J_1$ we have also integrated by parts. Notice that in the integration contour in $J_1$ can be shrinked to $\mathds{S}^1$, due to analytic properties of both $\omega_{\scaleto{<}{3pt}}$ and $\Omega_{\scaleto{>}{3pt}}$.
Moreover, using  Eq. \eqref{omegaomega2} we obtain
\begin{equation}
J_1=  -\int\limits_{\mathds{S}^1} \frac{dq}{2\pi i}\omega_{\scaleto{<}{3pt}}(q)\left(\omega'_{\scaleto{>}{3pt}}(q)-\sum_{k=1}^n \frac{1}{q-z_k}\right)=\int\limits_{\mathds{S}^1} \frac{dq}{2\pi i}\omega'_{\scaleto{<}{3pt}}(q)\omega_{\scaleto{>}{3pt}}(q)+\int\limits_{\mathds{S}^1} \frac{dq}{2\pi i}\sum_{k=1}^n \frac{\omega_{\scaleto{<}{3pt}}(q)}{q-z_k}.
\end{equation}
Now in the second integral, we can pull the contour to infinity and evaluate the residues at $q=z_k$ 
\begin{equation}
J_1=\int\limits_{\mathds{S}^1} \frac{dq}{2\pi i}\omega'_{\scaleto{<}{3pt}}(q)\omega_{\scaleto{>}{3pt}}(q)-\sum_{k=1}^n \omega_{\scaleto{<}{3pt}}(z_k).
\end{equation}
To evaluate $J_2$ we shrink the contour to the origin and obtain the following sum of the residues 
\begin{equation}
J_2=n\Omega_{\scaleto{>}{3pt}}(0)-\sum_{k=1}^n\Omega_{\scaleto{>}{3pt}}(z_k).
\end{equation}
The evaluation of $\Omega_{\scaleto{>}{3pt}}(z_k)$ should be understood as a limiting procedure from Eq. \eqref{omegaomega2}, namely 
\begin{equation}
	e^{\Omega_{\scaleto{>}{3pt}}(z_k)}=\lim\limits_{q\to z_k} \frac{e^{\omega_{\scaleto{>}{3pt}}(q)}}{\prod_{k=1}^n (z_k-q)}=\lim\limits_{q\to z_k} \frac{q^n(1+\theta(q))e^{\omega_{\scaleto{<}{3pt}}(q)} }{\prod_{j=1}^n (z_j-q)}=\frac{z_k^n \theta'(z_k)}{\prod_{j\neq k}^n (z_j-z_k)}e^{\omega_{\scaleto{<}{3pt}}(z_k)}.
\end{equation}
In the similar way we express  $\Omega_{\scaleto{>}{3pt}}(0)$ via $\omega_{\scaleto{>}{3pt}}(0)$
\begin{equation}
	\exp\left(\Omega_{\scaleto{>}{3pt}}(0)-\omega_{\scaleto{>}{3pt}}(0)\right)=\prod_{k=1}^n z_k^{-1}.
\end{equation}

Overall, we obtain 
\begin{equation}
	\exp\left(\int\limits_{\mathcal{C}} \frac{dq}{2\pi i}\Omega'_{\scaleto{<}{3pt}}(q)\Omega_{\scaleto{>}{3pt}}(q) \right)=  \exp\left(\int\limits_{\mathds{S}^1} \frac{dq}{2\pi i}\omega'_{\scaleto{<}{3pt}}(q)\omega_{\scaleto{>}{3pt}}(q) +n\omega_{\scaleto{>}{3pt}}(0)-2\sum\limits_{k=1}^n \omega_{\scaleto{<}{3pt}}(z_k)\right) \frac{\prod\limits_{k=1}^n\prod\limits_{j \neq k}^n(z_j - z_k)}{\prod\limits_{k=1}^n z_k^{2n} \theta'(z_k)},
\end{equation}
which results in Eq. \eqref{tauCreduced}

\section{Hartwig-Fisher asymptotic without effective form factors}\label{tauHF:app}
\subsection{Resolvent from the Riemann-Hilbert problem}
Here we derive relation \eqref{tauHF} without referring to the effective form factors series. We focus on the negative winding number for $\nu(q)$, $n=-\mathfrak{n}>0$. Instead, of deforming the contour to achieve zero winding as in our main approach we evaluate the determinant \eqref{tauEffDet} directly on $\mathds{S}^1$ and solve the corresponding Riemann-Hilbert problem (RHP) explicitly. Namely, we have to find the matrix-valued function $\chi(q)$ with the following properties

\begin{itemize}
    \item $\chi(q)$ is holomorphic and invertible outside $\mathds{S}^1$.
    \item $\chi(q) \to\mathds{1} $ as $q\to \infty$. 
    \item  For $q\in \mathds{S}^1$ we have a jump relation 
    \begin{equation}\label{jumpdef1}
\chi_{\scaleto{<}{3pt}}(q)  =\chi_{\scaleto{>}{3pt}}(q)  J,\qquad 
J =  \mathds{1} + \theta(q) |V(q)\rangle \langle V(q)|.
\end{equation}
\end{itemize}
The first two transformations are almost identical to those of Sec. \eqref{secS1}. 
Let us make transformations by the conjugation of the following matrices
	\begin{equation}\label{Phiintro}
		\chi = \Psi \left(\begin{array}{cc}
			e^{-\omega} & 0 \\
			0 & e^{\omega}
		\end{array}\right) \left(\begin{array}{cc}
			1 & 0 \\
			\varphi & 1
		\end{array}\right),
	\end{equation}
	where we introduced functions
	\begin{equation}\label{c3}
		\varphi(q) = \int\limits_{\mathds{S}^1} \frac{dk}{2\pi i} \frac{k^x}{k-q}\frac{\theta(k)}{1+\theta(k)},\qquad \omega(q) = \int\limits_{\mathds{S}^1} \frac{\nu_{\mathfrak{n}}(k)dk}{k-q}, 
	\end{equation}
with
	\begin{equation}\label{nudeltadefdiff}
		2\pi i\nu_{\mathfrak{n}}(k)= \ln[ (1+\theta(k))k^{-\delta}].
	\end{equation}
Notice that the index of $\nu_{\mathfrak{n}}(k)$ is zero on $\mathds{S}^1$. This definition differs by a constant from $\nu_{\mathfrak{n}}(k)$ defined in \eqref{nudeltadef}. 
The resulting jump for $\Psi$ reads 
        \begin{equation}
		\Psi_{\scaleto{>}{3pt}}^{-1}\Psi_{\scaleto{<}{3pt}}=J_{\Psi}=\left(\begin{array}{cc}
			q^{\mathfrak{n}} & e^{-\omega_{\scaleto{>}{3pt}} -\omega_{\scaleto{<}{3pt}}}\theta q^{-x} \\
			0 & q^{-\mathfrak{n}}
		\end{array}\right).
	\end{equation}
Further, we can make the following transformation to simplify the jump
\begin{equation}
	\Psi_{\scaleto{<}{3pt}}=\Phi_{\scaleto{<}{3pt}}\begin{pmatrix}
		1 & e^{-2\omega_{\scaleto{<}{3pt}}}q^{-x} \\
		0 & 1
	\end{pmatrix}, \qquad \Psi_{\scaleto{>}{3pt}}=\Phi_{\scaleto{>}{3pt}}
\end{equation}
the jump for this RHP $J_\Phi$ is presented as 
\begin{equation}
	\Phi_{\scaleto{>}{3pt}}^{-1}\Phi_{\scaleto{<}{3pt}}=J_\Phi=\left(\begin{array}{cc}
		q^{\mathfrak{n}} & -q^{-\mathfrak{n}}\mu(q) \\
		0 & q^{-\mathfrak{n}}
	\end{array}\right),
\end{equation}
where we introduced 
\begin{equation}\label{mudef}
	\mu(q)=e^{-\omega_{\scaleto{>}{3pt}} -\omega_{\scaleto{<}{3pt}}} q^{-x+\mathfrak{n}}.
\end{equation}
Up to this moment, all the transformations were holomorphic and invertible outside $\mathds{S}^1$. To proceed further, we define the new RHP as
\begin{equation}
	\Phi_{\scaleto{<}{3pt}}(q)=Y_{\scaleto{<}{3pt}}(q) q^{\mathfrak{n}\sigma_3}, \qquad  \Phi_{\scaleto{>}{3pt}}(q)=Y_{\scaleto{>}{3pt}}(q).
\end{equation}
The corresponding jump reads 
\begin{equation}\label{RHPy}
	Y_{\scaleto{>}{3pt}}^{-1}Y_{\scaleto{<}{3pt}}=J_Y=J_\Phi q^{-\mathfrak{n}\sigma_3}=
	\begin{pmatrix}
		1 & -\mu \\
		0 & 1
	\end{pmatrix}.
\end{equation}
Additionally to the jump condition, we changed normalization at infinity, which now has the form
        \begin{equation}\label{RHPynormaliz}
            Y_{\scaleto{<}{3pt}}(q)=\left(\mathds{1}+O(q^{-1})\right)q^{-\mathfrak{n}\sigma_3}, \qquad q\to\infty.
        \end{equation}
It is clear, that $Y(q)$ is not holomorphic at infinity. However, this RHP still can be solved using orthogonal polynomials as we consider $n=-\mathfrak{n}>0$ (see Remark 3.2 in \cite{Bertola2017}). We can formally define monic orthogonal polynomials with respect to the measure function $\mu(q)$ given by \eqref{mudef}
        \begin{equation}
            \int_{\mathds{S}^1} p_i(q)p_j(q)\mu(q)dq=h_i \delta_{ij}, \qquad h_i\neq 0.
        \end{equation}
The condition that $h_i\neq 0$ is quite nontrivial for generic measure function $\mu(q)$. It can be proven \cite{Chihara1978} that the following conditions are equivalent
\begin{itemize}
    \item Norms of polynomials $h_k\neq 0$ for $k=0, \ldots, n-1$;
    \item Determinants of Gram matrix restricted on the space of polynomials of degree less than $k$ are nonzero $\Delta_k\neq 0$ for $k=1, \ldots, n$;
        \begin{equation}
            \Delta_k=\det_{1\leq i,j \leq k} \mu_{i+j-2}, \qquad \mu_j=\int\limits_{\mathds{S}^1} k^j\mu(k)dk.
        \end{equation}
\end{itemize}
We will assume that our measure function $\mu(q)$ is such that $\Delta_k\neq 0$ for $k=1, \ldots, n$.
Then, the solution of the RHP \eqref{RHPy} reads
        \begin{equation}\label{RHPysol}
            Y(q)=\begin{pmatrix}
                \alpha(q) & A(q) \\
                \beta(q) & B(q)
            \end{pmatrix},
        \end{equation}
        where
        \begin{equation}
            \alpha(q)=p_n(q), \qquad \beta(q)=p_{n-1}(q)\cdot \left(-\frac{2\pi i}{h_{n-1}}\right)
        \end{equation}
        and
        \begin{equation}
            A(q)=\int_{\mathds{S}^1}\frac{dk}{2\pi i}\frac{\alpha(k)\mu(k)}{k-q}, \qquad B(q)=\int_{\mathds{S}^1}\frac{dk}{2\pi i}\frac{\beta(k)\mu(k)}{k-q}.
        \end{equation}
This can be readily checked using that $A_{\scaleto{>}{3pt}}-A_{\scaleto{<}{3pt}} = \alpha \mu$  and 
$B_{\scaleto{>}{3pt}}-B_{\scaleto{<}{3pt}} = \beta \mu$.
The non-trivial part is to satisfy the normalization conditions at $q\to \infty$ \eqref{RHPynormaliz}. First we notice that 
        \begin{equation}
            B_{\scaleto{<}{3pt}}(q)=\int_{|k|=1-0}\frac{dk}{2\pi i}\frac{\beta(k)\mu(k)}{k-q}=\sum_{j=1}^{\infty}q^{-j}\int_{\mathds{S}^1}k^{j-1}dk\frac{p_{n-1}(k)\mu(k)}{h_{n-1}}=q^{-n}+O(q^{-n-1}),
        \end{equation}
where we used $|k/q|<1$ and orthogonality condition for $p_n$. Similarly, we get $A_{\scaleto{<}{3pt}}(q)=O(q^{-n-1})$. Thus, we have
        \begin{equation}
            Y_{\scaleto{<}{3pt}}(q)q^{-n\sigma_3}=\mathds{1}+O(q^{-1}).
        \end{equation}
This finishes the proof for the solution of the RHP, which in the full form reads as
        \begin{equation}\label{rhsol1}
            \chi(q)=\begin{pmatrix}
                \alpha(q) & A(q) \\
                \beta(q) & B(q)
            \end{pmatrix} L(q)\left(\begin{array}{cc}
			e^{-\omega(q)} & 0 \\
			0 & e^{\omega(q)}
		\end{array}\right)
        \left(\begin{array}{cc}
			1 & 0 \\
			\varphi(q) & 1
		\end{array}\right),
        \end{equation}
where 
        \begin{equation}
            L_{\scaleto{<}{3pt}}(q)=q^{-n\sigma_3}\begin{pmatrix}
		1 & e^{-2\omega_{\scaleto{<}{3pt}}}q^{-x} \\
		0 & 1
	\end{pmatrix}, \qquad L_{\scaleto{>}{3pt}}(q)=1.
        \end{equation}
We can find resolvent using formulas \eqref{vectornotat} and \eqref{fv}.
\subsection{Variational formula}

Now let us apply the solution of the RHP from the previous subsection to find  $\tau^{\mathrm{eff}}[\nu]$ \eqref{tauEffDet}. 
To do this we consider the variation $\theta \to \theta+\delta \theta$ such that the winding number of $\nu_{\mathfrak{n}}$, defined in \eqref{nudeltadefdiff},  remains zero with respect to the contour $\mathds{S}^1$. 
The variation of the determinant can be computed in the same way as in \eqref{var2}
\begin{equation}
     \delta \ln \tau^{\mathrm{eff}}[\nu]
= 
     \int\limits_{\mathds{S}^1} dq \frac{\delta \theta(q)}{\theta(q)} R(q,q)+\int\limits_{|k|=1-0} \frac{dk}{2\pi i}\frac{\delta \theta(k)}{(1+\theta(k))^2}k^x M(k, k).
 \end{equation}
 Recall, that $M$ can be expressed with the solution of the RHP as follows (see Eq. \eqref{fromL1})
\begin{equation}
    M(k_1,k_2) = 
       \frac{\chi_{22}(k_2)\chi_{12}(k_1)-\chi_{22}(k_1)\chi_{12}(k_2)}{k_2-k_1}.
\end{equation}
Or using \eqref{RHPysol} and  \eqref{rhsol1}
\begin{equation}
    M(k_1,k_2)=\frac{Y_{22}(k_2)Y_{12}(k_1)-Y_{22}(k_1)Y_{12}(k_2)}{k_2-k_1}e^{\omega(k_1)+\omega(k_2)}.
\end{equation}
The derivation of this result is purely algebraic. It is convenient to introduce $2\times 2$ matrix $U(k_2,k_1)$ defined as
\begin{equation}
    Y^{-1}(k_2)Y(k_1)=\mathds{1}+(k_2-k_1)U(k_2,k_1),
\end{equation}
then we have
\begin{equation}
    M(k_1, k_2)=\left(U(k_2,k_1)\right)_{12}e^{\omega(k_1)+\omega(k_2)}.
\end{equation}
Note, that the matrix $U(k_2,k_1)$ has regular limit on diagonal $k_2=k_1$, which give us
\begin{equation}\label{Mdiag}
    M(k, k)=\left(U_{\scaleto{>}{3pt}}(k, k)\right)_{12}e^{2\omega_{\scaleto{>}{3pt}}(k)}, \qquad |k|<1.
\end{equation}
Let us also clarify the jump condition for the matrix $U$ on $\mathds{S}^1$. Denoting $J(q)\equiv J_Y(q) $ from Eq. \eqref{RHPy} we obtain 
\begin{multline}\label{jumpucond}
    U_{\scaleto{>}{3pt}}(p,q)=\frac{Y_{\scaleto{>}{3pt}}^{-1}(p)Y_{\scaleto{>}{3pt}}(q)-\mathds{1}}{p-q}=\frac{J(p)Y_{\scaleto{<}{3pt}}^{-1}(p)Y_{\scaleto{<}{3pt}}(q)J^{-1}(q)-\mathds{1}}{p-q}\\ =J(p)U_{\scaleto{<}{3pt}}(p,q)J^{-1}(q)+\frac{J(p)J^{-1}(q)-\mathds{1}}{p-q}.
\end{multline}
Let us express the resolvent \eqref{resolvent} via the matrix $U$. The numerator reads 
\begin{equation}
 \langle F(p)|F(q)\rangle = \langle V(p)|\chi^{-1}(p)\chi(q)|V(q)\rangle.
\end{equation}
We can express it in terms of $Y$. To do this, we will use \eqref{rhsol1} for $\chi_{\scaleto{<}{3pt}}(q)$. We have
\begin{equation}
   \langle F(p)|F(q)\rangle =\langle V_1(p)|Y^{-1}(p)Y(q)|V_1(q)\rangle=\langle V_1(p)|V_1(q)\rangle+(p-q)\langle V_1(p)|U_{\scaleto{<}{3pt}}(p,q)|V_1(q)\rangle,
\end{equation}
where
\begin{align}
    |V_1(q)\rangle=&L_{\scaleto{<}{3pt}}(q)e^{-\omega_{\scaleto{<}{3pt}}(p)\sigma_3}
        \left(\begin{array}{cc}
			1 & 0 \\
			\varphi_{\scaleto{<}{3pt}}(q) & 1
		\end{array}\right)|V(q)\rangle = \left(\begin{array}{c}
		   0\\
		     -e^{\omega_{\scaleto{<}{3pt}}(q)}q^{n+x}\text{v}_-(q) 
		\end{array}\right), \\
  \langle V_1(p)|=&\langle V(p)|
            \left(\begin{array}{cc}
			1 & 0 \\
			-\varphi_{\scaleto{<}{3pt}}(p) & 1
		\end{array}\right)
  e^{\omega_{\scaleto{<}{3pt}}(p)\sigma_3}L_{\scaleto{<}{3pt}}^{-1}(p) = (e^{\omega_{\scaleto{<}{3pt}}(p)}p^{n+x}\text{v}_-(p) , 0 ).
\end{align}
where we have used that $w_{\mathds{S}^1}(q)$ in the definition of $\text{v}_+(q)$ in Eq. \eqref{Vdef2} is nothing but $w_{\mathds{S}^1}=q^x + \varphi_{\scaleto{<}{3pt}}(q)$ (see \eqref{c3}). This way $
	\langle V_1(p)|V_1(q)\rangle=0
$ and the trace of the resolvent can be presented as
\begin{equation}
    R(q,q) = \theta(q) \langle V_1(q)|U_{\scaleto{<}{3pt}}(q,q)|V_1(q)\rangle = -e^{2\omega_{\scaleto{<}{3pt}}(q)}q^{2n+x} \left(U_{\scaleto{<}{3pt}}(q,q)\right)_{12}.
\end{equation}
Using formula \eqref{Mdiag}, we can rewrite variation of $\tau^{\mathrm{eff}}[\nu]$ as
\begin{equation}
    \delta \ln \tau^{\mathrm{eff}}[\nu]=\int_{\mathds{S}^1}dq \delta\theta(q) e^{2\omega_{\scaleto{<}{3pt}}(q)}q^{2n+x}\left(U_{\scaleto{>}{3pt}}(q,q)-U_{\scaleto{<}{3pt}}(q,q)\right)_{12}.
\end{equation}
The jump in the off-diagonal elements can be recast in the jump of the diagonal with the help of the condition  \eqref{jumpucond}
\begin{equation}
	\left(U_{\scaleto{>}{3pt}}(q,q)\right)_{12}-\left(U_{\scaleto{<}{3pt}}(q,q)\right)_{12}=\mu(q)\left[\left(U_{\scaleto{>}{3pt}}(q,q)\right)_{11}+\left(U_{\scaleto{<}{3pt}}(q,q)\right)_{11}\right]-\mu'(q),
\end{equation}
Moreover, taking into account that 
$e^{2\omega_{\scaleto{>}{3pt}}(k)}=e^{2\omega_{\scaleto{<}{3pt}}(k)}k^{2n}(1+\theta(k))^2$ 
together with 
\begin{equation}
	-k^{x+2n} e^{2\omega_{\scaleto{<}{3pt}}(k)}\mu'(k)=\left(\frac{n+x}{k}+\omega'_{\scaleto{>}{3pt}}(k)+\omega'_{\scaleto{<}{3pt}}(k)\right)\frac{1}{1+\theta(k)},
\end{equation}
and
\begin{equation}
	k^{x+2n} e^{2\omega_{\scaleto{<}{3pt}}(k)}\mu(k)=\frac{1}{1+\theta(k)},
\end{equation}
We present the variation as 
\begin{equation}\label{var3}
	\delta \ln \tau^{\mathrm{eff}}[\nu]= \int\limits_{\mathds{S}^1} \frac{dq}{2\pi i} \frac{\delta \theta(q)}{1+\theta(q)}\left(\omega'_{\scaleto{>}{3pt}}(q)+\omega'_{\scaleto{<}{3pt}}(q)+\frac{n+x}{q}+\left(U_{\scaleto{>}{3pt}}(q,q)\right)_{11}+\left(U_{\scaleto{<}{3pt}}(q,q)\right)_{11}\right).
\end{equation}
Plugging in the exact solution \eqref{RHPysol} we obtain
\begin{equation}
    (U(q,q))_{11}=-B(q)\alpha'(q)+\beta'(q)A(q)=\int_{\mathds{S}^1}\frac{\mu(k)dk}{2\pi i}\frac{\alpha(k)\beta'(q)-\alpha'(q)\beta(k)}{k-q}.
\end{equation}
Or equivalently
\begin{equation}
    (U(q,q))_{11}=\int_{\mathds{S}^1}\frac{\mu(k)dk}{2\pi i}\frac{\alpha(k)\beta'(k)-\alpha'(k)\beta(k)}{k-q},
\end{equation}
where we used the following identities
\begin{equation}
    \int_{\mathds{S}^1}\frac{\mu(k)dk}{2\pi i}\frac{\alpha(k)(\beta'(q)-\beta'(k))}{k-q}= \int_{\mathds{S}^1}\frac{\mu(k)dk}{2\pi i}\frac{(\alpha'(q)-\alpha'(k))\beta(k)}{k-q}=0.
\end{equation}
It is quite trivial and can be proven using the orthogonality condition. Indeed, we have
\begin{equation}
    \mathrm{deg}_k\frac{\beta'(q)-\beta'(k)}{k-q}=n-3, \qquad     \mathrm{deg}_k\frac{\alpha'(q)-\alpha'(k)}{k-q}=n-2,
\end{equation}
which is less than the degree of the polynomials $\alpha(k)$ and $\beta(k)$ correspondingly. 

Now let us consider how the variation of $\mu$ is connected to the variation of $\nu$. 
Using definition \eqref{mudef} we obtain 
\begin{equation}
    \delta\mu(k)=-\left(\delta\omega_{\scaleto{>}{3pt}}(k)+\delta\omega_{\scaleto{<}{3pt}}(k)\right)\mu(k)=\left(\int_{|q|=1+0}+\int_{|q|=1-0}\right) dq \frac{\delta\nu_{\mathfrak{n}}(q)\mu(k)}{k-q}.
\end{equation}
Eq. \eqref{nuTheta} relates variations of $\theta$ and $\nu$
\begin{equation}
    \delta \nu_{\mathfrak{n}}(q)=\frac{1}{2\pi i}\frac{\delta\theta(q)}{1+\theta(q)},
\end{equation}
which allows us to present
\begin{equation}
    \int\limits_{\mathds{S}^1} \frac{dq}{2\pi i} \frac{\delta \theta(q)}{1+\theta(q)}\left[\left(U_{\scaleto{>}{3pt}}(q,q)\right)_{11}+\left(U_{\scaleto{<}{3pt}}(q,q)\right)_{11}\right]=\int\limits_{\mathds{S}^1} \frac{dk}{2\pi i} \delta\mu(k)\left(\alpha(k)\beta'(k)-\alpha'(k)\beta(k)\right).
\end{equation}
The last expression can be found from the variation of the Hankel (Toeplitz) determinant
\begin{equation}\label{idcristof}
    \delta\ln\det_{1\leq i,j \leq n} M_{ij}=\int_{\mathds{S}^1}  \frac{dk}{2\pi i} \delta\mu(k)\left(\alpha(k)\beta'(k)-\alpha'(k)\beta(k)\right),
\end{equation}
where the matrix $M_{ij} \equiv \mu_{n-1+i-j}$ is constructed from the moments of $\mu(q)$.
To prove \eqref{idcristof} we make use of the Christoffel--Darboux kernel defined as
\begin{equation}
    \mathcal{K}_{n-1}(q,k):=\sum_{j=0}^{n-1} \frac{p_j(q)p_j(k)}{h_j}=\sum_{i,j=1}^n q^{n-i}k^{j-1}A_{ij}.
\end{equation} 
That it is the kernel of orthogonal projection to the space of polynomials of $\mathrm{deg}\le n-1$
which can be explicitly expressed as 
\begin{equation}
       \sum_{j=0}^{n-1}\frac{p_j(q)p_{j}(k)}{h_j}=\frac{1}{h_{n-1}}\frac{p_{n}(k)p_{n-1}(q)-p_{n}(q)p_{n-1}(k)}{k-q} = \frac{\alpha(q)\beta(k)-\alpha(k)\beta(q)}{2\pi i (k-q)}.
\end{equation}
In the $k\to q$ limit this expression allow us to conclude 
\begin{equation}
    \frac{1}{2\pi i}\left(\alpha(k)\beta'(k)-\alpha'(k)\beta(k)\right)=\mathcal{K}_{n-1}(k,k)=\sum_{i,j=1}^n k^{n-i+j-1}A_{ij}.
\end{equation}
The matrix $A_{ij}$ is, in fact, the inverse of $M_{ij}$:  $(M^{-1})_{ij}=A_{ij}$ (see for example \cite{BarrySimons}). This way, the variation \eqref{idcristof} reads 
\begin{equation}
    \int_{\mathds{S}^1} \frac{dk}{2\pi i} \delta\mu(k)\left(\alpha(k)\beta'(k)-\alpha'(k)\beta(k)\right)=\sum_{i,j=1}^n A_{ij} \delta\mu_{n-i+j-1}=\sum_{i,j=1}^n (M^{-1})_{ij}\delta M_{ji}=\delta \ln\det_{1\leq i,j \leq n} M_{ij}.
\end{equation}
Let us return to the variation formula \eqref{var3}. The first three terms can be integrated into 
\begin{equation}
    \int_{\mathds{S}^1} \frac{dq}{2\pi i} \frac{\delta \theta(q)}{1+\theta(q)}\left(\omega'_{\scaleto{>}{3pt}}(q)+\omega'_{\scaleto{<}{3pt}}(q)+\frac{n+x}{q}\right) = \delta S[\nu_{\mathfrak{n}}]  
\end{equation}
where 
\begin{equation}
    S[\nu_{\mathfrak{n}}]= (x+n)\int_{\mathds{S}^1}\frac{dq}{q}\nu_{\mathfrak{n}} (q)-\frac{1}{2}\int_{\mathds{S}^1}\int_{\mathds{S}^1} dqdp\left(\frac{\nu_{\mathfrak{n}}(q)-\nu_{\mathfrak{n}}(p)}{q-p}\right)^2.
\end{equation}
The last two terms can be integrated into
\begin{equation}
    \int\limits_{\mathds{S}^1} \frac{dq}{2\pi i} \frac{\delta \theta(q)}{1+\theta(q)}\left[\left(U_{\scaleto{>}{3pt}}(q,q)\right)_{11}+\left(U_{\scaleto{<}{3pt}}(q,q)\right)_{11}\right] \quad \Rightarrow \quad \log\det_{1\leq i,j \leq n} \mu_{n-1+i-j}.
\end{equation}
Overall, we conclude that our original tau function has the same variation as asymptotics in the Hartwig--Fisher formula
\begin{equation}\label{vareq}
    \delta\log\tau^{\mathrm{eff}}[\nu]=\delta\log\tau_{\mathrm{HF}}[\nu],
\end{equation}
where
\begin{equation}
    \tau_{\mathrm{HF}}[\nu]\equiv (-1)^{n x}\det_{1\leq i,j \leq n} y_{\mathfrak{n}}(x+i-j)\times\exp\left(S[\nu_{\mathfrak{n}}]\right),
\end{equation}
with
\begin{equation}
    y_{\mathfrak{n}}(x+s)=\frac{1}{2\pi i}\int_{\mathds{S}^1} \frac{dk}{k} k^{-x-s} e^{-2\pi i\nu_{\mathfrak{n}}(k)}e^{-2\omega_{\scaleto{<}{3pt}}(k)}=\frac{1}{2\pi i}\int_{\mathds{S}^1} \frac{dk}{k} k^{n-s}\mu(k)=\frac{\mu_{n-1-s}}{2\pi i}.
\end{equation}
Note, that from equality of variations \eqref{vareq} follows that these tau functions might differ by an overall constant. However, in the main text, we show by comparing asymptotics that they are equal. Note, that the factor $(-1)^{nx}$ in the definition of $\tau_{\mathrm{HF}}$ can be eaten up by redefinition $\nu_{\mathfrak{n}}$ as in the main text \eqref{nudeltadef}.


 \bibliographystyle{elsarticle-num} 
 \bibliography{cas-refs}





\end{document}